\documentclass[sn-mathphys,Numbered]{sn-jnl}

\usepackage{graphicx}%
\usepackage{multirow}%
\usepackage{amsmath,amssymb,amsfonts}%
\usepackage{amsthm}%
\usepackage{mathrsfs}%
\usepackage[title]{appendix}%
\usepackage{xcolor}%
\usepackage{textcomp}%
\usepackage{manyfoot}%
\usepackage{booktabs}%
\usepackage{listings}%

\begin{document}

\title[Series expansion of potential]{Potential series expansion method applied in Analytical Modeling of Gravitational field of Irregularly Shaped Celestial Bodies
}


\author*[1,2]{\fnm{Marcelo L.} \sur{Mota}}\email{prof.mlmota@ifsp.edu.br}

  \author[1]{\fnm{Safwan} \sur{Aljbaae}}\email{safwan.aljbaae@gmail.com}
  \equalcont{These authors contributed equally to this work.}

  \author[1]{\fnm{Antonio F. B. A.} \sur{Prado}}\email{antonio.prado@inpe.br}
  \equalcont{These authors contributed equally to this work.}

  \affil*[1]{\orgdiv{Post Graduate Division}, \orgname{INPE}, \orgaddress{\street{Av. dos Astronautas, 1.758 - Jardim da Granja}, \city{S\~{a}o Jose dos Campos}, \postcode{12227-010}, \state{S\~{a}o Paulo}, \country{Brazil}}}
  \affil[2]{\orgdiv{Federal Institute of S\~{a}o Paulo}, \orgname{IFSP}, \orgaddress{\street{ Avenida Thereza Ana Cecon Breda, s/n - Vila S\~{a}o Pedro}, \city{Hortol\^{a}ndia}, \postcode{13183-250}, \state{SP}, \country{Brazil}}}


\abstract{
This study aims to establish an analytical model that reproduces the gravitational field around non-spherical bodies with constant density. Due to the non-spherical geometry of such bodies, their gravitational potential is disturbed relative to a central field. By considering the body as a polyhedron and decomposing it into tetrahedral elements, we use the Series Potential Expansion Method (PSEM) to approximate the total potential by summing the potentials of each tetrahedron. While this model does not offer higher accuracy than the classical polyhedral approach, it achieves relative errors below 0.1\% for points outside the body when developed to higher orders (e.g., orders 11 and 12), and significantly reduces execution time. To validate this approach, we apply our model to asteroids (87) Sylvia, (101955) Bennu, (99942) Apophis, and (25143) Itokawa. We determine equilibrium points, analyze stability, investigate zero-velocity planes, and calculate the relative errors between the gravitational field modeled by PSEM and the results obtained using both the classical polyhedral method by Tsoulis and Petrovic and the mass concentration method. Our results highlight the computational efficiency of PSEM in modeling the gravitational potential of irregularly shaped bodies. This efficiency stems from expressing the gravitational potential through a homogeneous analytical function that is easy to manipulate algebraically, enabling explicit determination of the acceleration vector. Our model provides a robust framework for more complex analyses, such as studying periodic orbits around non-spherical celestial bodies, assessing their stability, and planning the smooth landing trajectories of spacecraft.

}




\maketitle

\section{Introduction}\label{sec1}
    
Analyzing the dynamics of objects close to asteroids has garnered increasing interest due to its potential to provide crucial information about the early stages of the formation and evolution of our Solar System. Understanding the behavior of spacecraft and other objects near asteroids is not only scientifically fascinating, but also important for space exploration and planetary defense. However, studying these dynamics presents a significant challenge due to the non-central gravitational force. A spacecraft in orbit around an asteroid will be subjected to various disturbing forces, including the irregular gravitational field generated by the body's non-sphericity, solar radiation pressure (SRP), and the gravitational effects of the Sun, all of which significantly alter its orbital elements and thus change its trajectory over time. Therefore, the key challenge in mission design is to develop a mathematical model that accurately reflects the gravitational field outside the asteroid.\

Previous studies have explored different methodologies for modeling the gravitational field around irregular celestial bodies, providing valuable insights into the intricate challenges posed by non-spherical bodies.\

The spherical harmonic method aims to model the gravitational potential using a harmonic series, as developed by \citet{Macmillan_1930}, \citet{1966tsga.book.....K}, \citet{1967phge.book.....H}, \citet{1994CeMDA..60..331B}, \citet{2019AIAAJ..57.4291R} and \citet{2000DPS....32.6529C}. The coefficients are calculated using volume integrals, which can be reduced to surface integrals under the assumption of constant density. This approach offers a convergent series gravitational model, mainly suitable for points outside the circumscribed Brillouin sphere. However, the method approximates the gravitational field through a finite series, where truncation error increases near the convergence radius, so requiring additional terms to maintain accuracy. Furthermore, this method may diverge within the circumscribed sphere, particularly for bodies with non-spherical geometry or concave shapes.

The classical Mascon approach models gravitational fields of irregular bodies by placing point masses that replicate the object's mass distribution, as proposed by \citet{1996Icar..120..140G}. This method is known for its simplicity and computational speed. Building on this concept, \citet{chanut_2015, aljbaae_2017, aljbaae_2021} introduced enhancements using a polyhedral source, achieving reduced execution time while maintaining satisfactory accuracy. Additionally, \citet{chanut_2015, aljbaae_2017, aljbaae_2021} demonstrated that this method could effectively address limitations of the polyhedral approach by incorporating layered density structures within the asteroid. In our previous studies, we further explored this method by testing configurations with one, four, and eight point masses per tetrahedron. While these configurations offered reasonable results, they proved insufficiently accurate near the surface of irregular bodies. To strike a balance between precision and efficiency, we adopted a configuration with 20 point masses per tetrahedron in this work. This choice enables us to demonstrate the accuracy of the Series Potential Expansion Method (PSEM) while achieving a computational time reduction of over 80\% compared to the classical polyhedral method. Although higher point mass configurations could enhance accuracy further, they would also significantly increase execution time, which would detract from the computational efficiency we aim to emphasize.

The Polyhedron approach, developed by \citet{Werner_1994}, models the gravitational field around irregular bodies comprehensively. The body is conceptualized as a polyhedron, with its surface divided into disjoint triangular regions that connect to the center of mass, forming tetrahedrons. The method involves intricate calculations to accurately evaluate gravitational potential and gradients. While it is one of the most accurate methods, it requires substantial computational resources due to the significant number of triangular faces. \citet{tsoulis_2001} analyzed potential field singularities arising from this approach.

In this study, we introduce the Potential Series Expansion Method (PSEM) developed in \citet{Mota_2017, Mota_2019, Mota_2023}, which combines potential series expansion with asteroid decomposition into tetrahedral elements. This method assumes homogeneous density and constant rotation speed, providing an analytical representation of the gravitational potential, while reducing computational expenses for orbital simulations.

Section 2 presents the series potential expansion method. Section 3 provides details of the numerical simulations and script sequence. In Section 4 we apply our approach to asteroids (87) Sylvia, (101955) Bennu, (99942) Apophis and (25143) Itokawa, comparing our results with the classical polyhedral method. Finally, Section 5 offers discussion and conclusions.

\section{Potential series expansion method}\label{Method_series_expansion}

    Fundamentally based on the studies made by \citet{Kellogg_1954} and \citet{Macmillan_1930}, and incorporating the method of polyhedron decomposition into tetrahedra, as developed by \citet{Werner_1994}, this study presents a methodology to approximate the potential of a body with asymmetric mass distribution. It concludes that the potential of a polyhedron is obtained by summing the potentials of each generated tetrahedron. Consider a homogeneous solid with a non-spherical geometry Q, whose center of mass is at the origin of the Cartesian coordinate system \(\xi, \eta,\zeta \). To apply the polyhedral method developed by \citet{Werner_1996}, it's necessary to model the shape of this solid as a polyhedron by summing the tetrahedral elements. This is done by connecting the vertices of triangular bases on its surface to its center of mass, as shown in Figure 1.

    \begin{figure}
        \centering
        \includegraphics[width=0.50\textwidth]{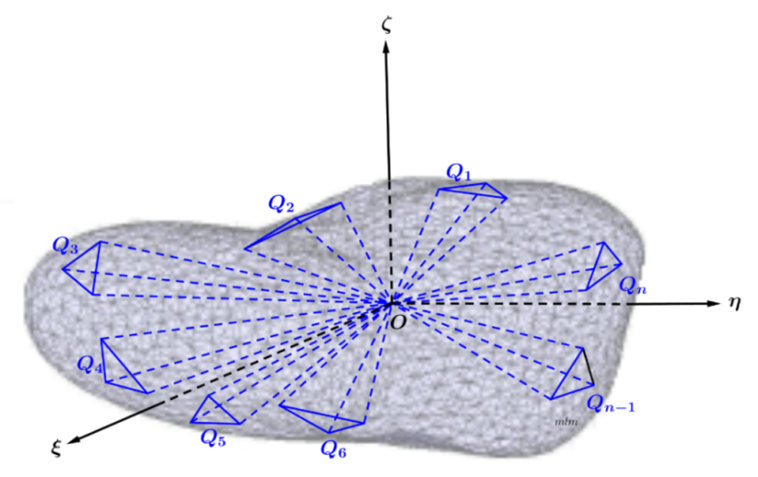}
        \caption{Homogeneous solid Q with non-spherical geometry decomposed into tetrahedra.}
        \label{tetrahedra}
    \end{figure}

    Consider \({Q_k}\) as the generic tetrahedron of mass \({M_k}\), formed by the vertices \({V_{{1_k}}}\), \({V_{{2_k}}}\), \({V_{{3_k}}}\), and O, with the latter positioned at the origin of the Cartesian coordinate system \(\xi,\eta,\zeta \), as illustrated in Figure 2.

    \begin{figure}
        \centering
        \includegraphics[width=0.50\textwidth]{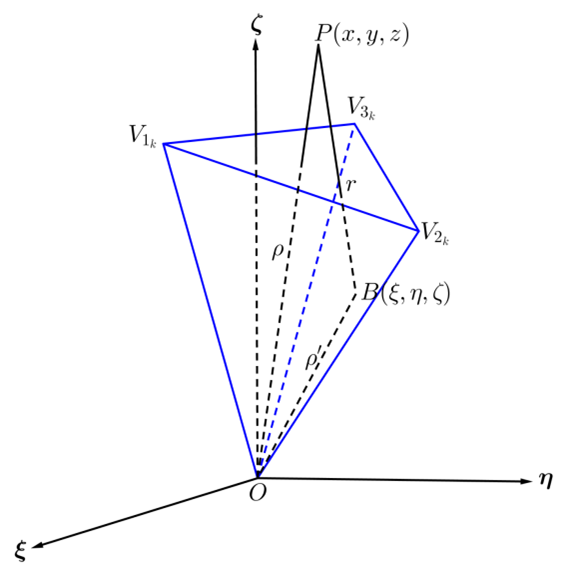}
        \caption{Tetrahedron \({Q_k}\) with vertices  \({V_{{1_k}}}\), \({V_{{2_k}}}\), \({V_{{3_k}}}\), and O.}
        \label{Tetrahedron}
    \end{figure}

    Consider a unit mass at a point \(P\left( {x,y,z} \right)\) and an element of mass $dM$ at any point in the tetrahedron \({Q_k}\), denoted as \(B\left( {\xi ,\,\eta ,\,\zeta } \right)\). Defining \(\gamma \) as the angle between the vectors $\overrightarrow{\text{OP}}$ and $\overrightarrow{\text{OB}}$, from Figure 2, we have the Equation (1):
    \begin{eqnarray}\label{eq_1}
      \frac{1}{r} = \frac{1}{{\sqrt {{\rho ^2} + {{\rho '}^2} - 2\rho \rho '\cos \gamma } }},
    \end{eqnarray}

    and placing \({\rho ^2}\) in evidence, results Equation (2):
    \begin{eqnarray}\label{eq_2}
      \frac{1}{r} = \frac{1}{{\rho \sqrt {1 - 2\frac{{\rho '}}{\rho }\cos \gamma  + \frac{{{{\rho '}^2}}}{{{\rho ^2}}}} }}
    \end{eqnarray}

    Defining \(\chi  = {{\rho '} \mathord{\left/{\vphantom {{\rho '} \rho }} \right.
    \kern-\nulldelimiterspace} \rho }\) and \(u = \cos \gamma \), then Eq. (2) will become Equation (3):
    \begin{eqnarray}\label{eq_3}
      \frac{1}{r} = \frac{1}{{\rho \sqrt {1 - 2\chi u + {\chi ^2}} }}
    \end{eqnarray}

    Developing \[{\left( {\sqrt {1 - 2u\chi  + {\chi ^2}} } \right)^{ - 1}}\] in a power series of \(\chi \) convergent, we obtain Equation (4):
    \begin{eqnarray}\label{eq_4}
      \frac{1}{{\,\sqrt {1 - 2u\chi  + {\chi ^2}} }} = \sum_{i=0}^{m}{P_i}\left( u \right)\chi^i  
    \end{eqnarray}
    as long as \(|\chi |\, < \sqrt 2 - 1\). The terms \({P_o}\left( u \right),\,{P_1}\left( u \right),\,{P_2}\left( u \right),...\) represent the associated Legendre polynomials. Thus, Equation 4 can be written as a series of negative powers of the variable \(\rho \) in Equation (5):
    \begin{eqnarray}\label{eq_5}
      \frac{1}{r} = {P_o}\left( u \right)\frac{1}{\rho } + {P_1}\left( u \right)\frac{{\rho '}}{{{\rho ^2}}} + {P_2}\left( u \right)\frac{{{{\rho '}^2}}}{{{\rho ^3}}} + ...\,
    \end{eqnarray}
    This holds if \({\rho } < \sqrt 2  - 1\), but it also converges when \( - 1 \le u \le 1\) and \({{\rho '} \mathord{\left/ {\vphantom {{\rho '} \rho }} \right. \kern-\nulldelimiterspace} \rho } < 1\). If $d$ is the maximum distance between a mass distribution point\footnote{A 'mass distribution point' refers to any point within the polyhedral body where the mass is considered to contribute to the gravitational potential during the integration process over the tetrahedral elements.} and the origin, then the series given by Equation (5) will converge uniformly for any point $P$ outside a sphere centered at the origin with radius \(\rho  = d.\tau \), where \(\tau  > 1\) (\(d.\tau \) represents the product of $d$ and $\tau$). In this case, the sphere corresponds to the Brillouin sphere, the smallest sphere enclosing the entire mass of the body. Inside this sphere, similarly to the spherical harmonics method, the series may diverge. However, outside the Brillouin sphere, the method converges uniformly, ensuring reliable approximations of the gravitational field.\\

    Thus, Equation (5) can be multiplied by a continuous (or piecewise continuous) density and integrated term by term over the solid volume \({Q_k}\). This results in an expansion of the potential \({U^{\left( k \right)}}\) for the mass distribution of \({Q_k}\), which is uniformly convergent outside every sphere containing the distribution, given in Equation (6):

    \begin{eqnarray}\label{eq_6}
      {{U}^{\left( k \right)}}=\iiint\limits_{{{Q}_{k}}}{G\sigma \frac{1}{r}dV}
    \end{eqnarray}
where \(dV = d\xi \,d\eta \,d\zeta \). Assuming that the density of the solid, \(\sigma  = {M \mathord{\left/{\vphantom {M V}} \right. \kern-\nulldelimiterspace} V}\), is constant, and integrating term by term in the right member of Equation (6) yields in equation (7):
    \begin{eqnarray}\label{eq_7}
     {{U}^{\left( k \right)}}=G\sigma \sum\limits_{i=0}^{\infty }{\iiint\limits_{{{Q}_{k}}}{{{P}_{i}}\left( u \right)\frac{{{{{\rho }'}}^{i}}}{{{\rho }^{i+1}}}dV}}
    \end{eqnarray}
    Equation (7) should be truncated to obtain an approximate potential for the tetrahedron \({Q_k}\), as shown in Equation (8):

    \begin{eqnarray}\label{eq_8}
    {{U}^{\left( k \right)}}=G\sigma \sum\limits_{i=0}^{m}{\iiint\limits_{{{Q}_{k}}}{{{P}_{i}}\left( u \right)\frac{{{{{\rho }'}}^{i}}{{\rho }^{i}}}{{{\rho }^{2i+1}}}dV}}+{{\varepsilon }_{k}}
    \end{eqnarray}
 where \({\varepsilon _k}\) is the truncation error. As Equation 8's series is uniformly convergent within the adopted interval, the larger the value of $m$, the smaller \({\varepsilon _k}\), balancing computational cost against precision. Furthermore, from Figure 2, $\cos \gamma =u$, $\rho =\sqrt{{{x}^{2}}+{{y}^{2}}+{{z}^{2}}}$ and ${\rho }'=\sqrt{{{\xi }^{2}}+{{\eta }^{2}}+{{\zeta }^{2}}}$, we have Equation (9).
    \begin{eqnarray}\label{eq_9}
    \,\,\,\,u\rho \rho ' = \xi x + \eta y + \zeta z,
    \end{eqnarray}
    Thus, substituting the Legendre polynomials, \(\rho \), \(\rho'\), and considering Equation (9), each integrand in Equation 8 is a homogeneous polynomial function of \(\xi \), \(\eta \), and \(\zeta \), given by Equation (10):
    \begin{eqnarray}\label{eq_10}
    f\left( \xi ,\eta ,\varsigma  \right)=\sum\limits_{{{n}_{1}},{{n}_{2}},{{n}_{3}}}^{{}}{{{\xi }^{{{n}_{1}}}}{{\eta }^{{{n}_{2}}}}{{\zeta }^{{{n}_{3}}}}}.
    \end{eqnarray}

    In Equation (10), the indices \(n_1, n_2, \), and \(n_3\) denote the powers of the variables \(\xi \), \(\eta \), and \(\zeta \) in the polynomial expansion. The summation is performed over all combinations of these indices, up to the specified order of the expansion, allowing for an accurate approximation of the gravitational potential.\\


    Thus, determining the potential is linked to the integral calculations in Equation (8), which can be simplified by using the method developed by \citet{Lien_1984}. This symbolic method is fundamental to our approach, as it establishes a linear transformation (isometry) between the points of a generic tetrahedron and the points of a rectangular tetrahedron, both with one vertex at the origin of their respective coordinate systems. This transformation resolves the primary difficulty in calculating triple integrals over a generic tetrahedron: determining the integration limits. By transforming to a rectangular tetrahedron, these limits are well-defined and significantly simplified.

    The method further leverages the Beta and Gamma functions and their relationships to streamline the computation of the triple integral. The result of this integral is then multiplied by the Jacobian determinant of the transformation matrix to map the solution back to the original coordinate system. This provides the value of the potential generated by the generic tetrahedron. By summing the contributions from all tetrahedra, the total gravitational potential of the polyhedron is determined.

    We note that the coordinate system is fixed in the polyhedron (asteroid), with the origin located at the center of mass and the axes aligned with the principal axes of inertia. This configuration ensures consistency in our calculations across all tetrahedra that compose the polyhedron.

    This combination of geometric and mathematical simplifications is a cornerstone of the Series Potential Expansion Method (PSEM) and is critical to its efficiency and accuracy. To calculate the integrals, we focus on the integral in Equation 11. For additional details on the symbolic method and its applications, see \citet{Mota_2017}.

    \begin{eqnarray}\label{eq_11}
      I=\iiint\limits_{Q}{{{\xi }^{{{n}_{1}}}}{{\eta }^{{{n}_{2}}}}{{\zeta }^{{{n}_{3}}}}d\xi d\eta d\zeta }.
    \end{eqnarray}
    Assuming the tetrahedron vertices in Figure 3 are \(O\left( {0,0,0} \right)\), \({V_1}_k\left( {{\xi }_{{{1}_{k}}},{\eta }_{{{1}_{k}}},{\zeta }_{{{1}_{k}}}} \right)\), \({V_2}_k\left( {{\xi }_{{{2}_{k}}},{\eta }_{{{2}_{k}}},{\zeta }_{{{2}_{k}}}} \right)\) and \({V_3}_k\left( {{\xi }_{{{3}_{k}}},{\eta }_{{{3}_{k}}},{\zeta }_{{{3}_{k}}}} \right)\), the linear transformation $T$ is expressed by Equation (12).
    \begin{eqnarray}\label{eq_12}
        T=\left( \begin{matrix}
    {{\xi }_{{{1}_{k}}}} & {{\xi }_{{{2}_{k}}}} & {{\xi }_{{{3}_{k}}}}  \\
   {{\eta }_{{{1}_{k}}}} & {{\eta }_{{{2}_{k}}}} & {{\eta }_{{{3}_{k}}}}  \\
   {{\zeta }_{{{1}_{k}}}} & {{\zeta }_{{{2}_{k}}}} & {{\zeta }_{{{3}_{k}}}}  \\
    \end{matrix} \right).
    \end{eqnarray}

    This establishes an isometry between the coordinate system \(\left( {\xi,\eta,\zeta} \right)\) and the new system \(\left( {X,Y,Z} \right)\), as shown in Equation (13).
    \begin{eqnarray}\label{eq_13}
        \left( \begin{matrix}
    \xi   \\
    \eta   \\
    \zeta   \\
    \end{matrix} \right)=\left( \begin{matrix}
    {{\xi }_{{{1}_{k}}}} & {{\xi }_{{{2}_{k}}}} & {{\xi }_{{{3}_{k}}}}  \\
   {{\eta }_{{{1}_{k}}}} & {{\eta }_{{{2}_{k}}}} & {{\eta }_{{{3}_{k}}}}  \\
   {{\zeta }_{{{1}_{k}}}} & {{\zeta }_{{{2}_{k}}}} & {{\zeta }_{{{3}_{k}}}}  \\
    \end{matrix} \right)\left( \begin{matrix}
    X  \\
    Y  \\
    Z  \\
    \end{matrix} \right)
    ,
    \end{eqnarray}

    This transformation changes the initial tetrahedron to a right-angled tetrahedron with vertices  \(O'\left( {0,0,0} \right)\), \({V'_1}_k\left( {1,0,0} \right)\), \({V'_2}_k\left( {0,1,0} \right)\) and \({V'_3}_k\left( {0,0,1} \right)\), as shown in Figure 3.

    \begin{figure}
        \centering
        \includegraphics[width=0.50\textwidth]{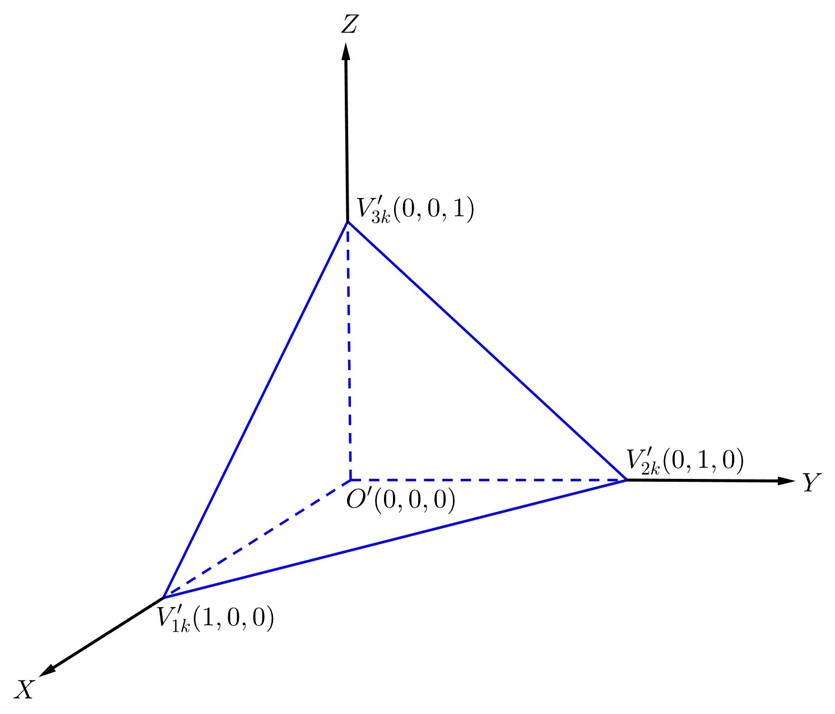}
        \caption{Tetrahedron \({W}\) with vertices  \({V'_{{1_k}}}\), \({V'_{{2_k}}}\), \({V'_{{3_k}}}\), and O.}
        \label{Tetrahedron}
    \end{figure}

    According to the transformation given by Equation 12, the integral in Equation (11) can be rewritten in Equation (14).

    \begin{eqnarray}\label{eq_14}
     I=||T||\iiint\limits_{W}{{{{\tilde{X}}}^{{{n}_{1}}}}{{{\tilde{Y}}}^{{{n}_{2}}}}{{{\tilde{Z}}}^{{{n}_{3}}}}dX\,dY\,dZ}
    \end{eqnarray}
where $\tilde{X}={{\xi }_{1k}}X+{{\xi }_{2k}}Y+{{\xi }_{3k}}Z$, $\tilde{Y}={{\eta }_{1k}}X+{{\eta }_{2k}}Y+
    {{\eta }_{3k}}Z$, $\tilde{Z}={{\zeta }_{1k}}X+{{\zeta }_{2k}}Y+{{\zeta }_{3k}}Z$, while $||T||$ is the
    Jacobian determinant of the transformation matrix. Evaluating the integral in Equation 11 over the $W$ 
    tetrahedron yields Equation (15).

   \begin{eqnarray}\label{eq_15}
    \int\limits_{0}^{1}{\int\limits_{0}^{1-Y }{\int\limits_{0}^{1-Y -Z }{{{X }^{{{n}_{_{1}}}}}{{Y }^{{{n}_{_{2}}}}}{{Z }^{{{n}_{_{3}}}}}dX dY dZ =\frac{{{n}_{1}}\text{! }{{n}_{2}}\text{! }{{n}_{3}}\text{!}}{\left( {{n}_{1}}+{{n}_{2}}+{{n}_{3}}+3 \right)\text{!}}}}} \
    \end{eqnarray}

    To clarify the transition from Equation (14) to Equation (15), the symbolic method developed by Lien and Kajiya is applied. This involves establishing an isometry between the generic tetrahedron and the rectangular tetrahedron, which simplifies the limits of integration. The integral over the rectangular tetrahedron in Equation (15) is evaluated by leveraging the relationship between the Beta and Gamma functions  (as detailed in Appendix \ref{appendice_a}), which simplifies the computation of the triple integral. The result is then multiplied by the Jacobian of the transformation, mapping the solution back to the original coordinate system.\\

    Thus, Equation (14) represents the integral applied in the transformed coordinate system, while Equation (15) shows the simplified integral evaluated over the rectangular tetrahedron in the original coordinate system. This process ensures a smooth transition between the two equations and highlights the efficiency of the method.\\

    Thus, the potential of the non-spherical geometry Q relative to the unit mass point \(P\left( {x,y,z} \right)\) is given by Equation (16), where \(\varepsilon  = \sum\limits_{k = 1}^n {{\varepsilon _k}} \).

    \begin{eqnarray}\label{eq_16}
      {{U}^{\left( m \right)}}=G\sigma \sum\limits_{k=1}^{n}{\sum\limits_{i=0}^{m}{\iiint\limits_{{{Q}_{k}}}{{{P}_{i}}\left( u \right)\frac{{{{{\rho }'}}^{i}}{{\rho }^{i}}}{{{\rho }^{2i+1}}}dV}}} +\varepsilon
    \end{eqnarray}

    Defining \({U_i}\), \(i = 0,1,2,...,m\), as the potential of degree $i$, satisfying Equation (17)

    \begin{eqnarray}\label{eq_17}
     {{U}_{i}}=G\sigma \sum\limits_{k=1}^{n}{\iiint\limits_{{{Q}_{k}}}{{{P}_{i}}\left( u \right)\frac{{{{{\rho }'}}^{i}}{{\rho }^{i}}}{{{\rho }^{2i+1}}}dV}}
    \end{eqnarray}

    we conclude that the gravitational field model of the homogeneous solid Q, using the potential series expansion method, results in Equation (18).
    \begin{eqnarray}\label{eq_18}
      U = {U_0} + {U_1} + {U_2} + ... + {U_m}
    \end{eqnarray}
    Note that \({U_0}\) corresponds to the Keplerian potential, while the sum of the other terms represents the perturbation due to the non-central field. We will use the notation:
    \begin{eqnarray}\label{eq_21}
    \text{Pot }n=\sum\limits_{i=0}^{n}{{{U}_{i}}}
    \end{eqnarray}

\section{Equations of motion and equilibrium points}\label{Equations of motion and equilibrium points}

In this work, only the disturbance generated by the gravitational field was taken into account, leaving other 
disturbances, such as solar radiation pressure and the effects of solar gravitation for a later study.

\subsection{Equations of motion}\label{Equations of motion}

We adopted the coordinate system fixed to the asteroid for all equations. Therefore, to obtain the potentials 
of the asteroids studied in this work, the PSEM series potential expansion method was used, associated with 
the decomposition of the asteroid into tetrahedral elements (\citet{Mota_2017, Mota_2019, Mota_2023}), that is, 
we used equation (17).\\
As mentioned previously, the asteroid studied has a rotational speed only relative to the z axis. Assuming this uniform velocity, the equations of motion of a particle orbiting around the asteroid (\citet{greenwood1988principles} and \citet{1996Icar..121...67S})
 are expressed in the form of Equation (20).
 \begin{eqnarray}\label{eq_20}
     \mathbf{\ddot{r}}+2\left( \omega \times \mathbf{\dot{r}} \right)+\omega \times \left( \omega \times \mathbf{r} \right)=-\frac{\partial \text{$U$}}{\partial \mathbf{r}}
    \end{eqnarray}\
with \textbf{r} being the position vector of the particle, \textbf{\(\dot{r}\)}  the first and \textbf{\(\ddot{r}\)} the second temporal derivatives relative to the system fixed to the asteroid, \textbf{\(\omega\)} the angular velocity vector of the asteroid relative to the inertial system and $U$ being the gravitational potential of the asteroid, given by Equation (18). We can define the effective potential function, using Equation (21).
\begin{eqnarray}\label{eq_21}
     V\left( \text{r} \right)=-\frac{1}{2}\left( \mathbf{\omega }\times \text{r} \right)\cdot \left( \mathbf{\omega }\times \text{r} \right)+U\left( \text{r} \right)\
    \end{eqnarray}\
It makes possible to rewrite Equation (20) in Equation (22)
\begin{eqnarray}\label{eq_23}
     \mathbf{\ddot{r}}+2\left( \omega \times \mathbf{\dot{r}} \right)+\frac{\partial V}{\partial \mathbf{r}}=0\
    \end{eqnarray}\
In this work, we consider smaller bodies in uniform rotation, and we assume that $\text{ }\!\!\omega\!\!\text{ }=\omega \,{{\textbf{e}}_{\text{z}}}$, where ${\textbf{e}}_{\text{z}}$ is the unit vector referring to the $z$ axis, simplifying Equation (22). 

\subsection{Determination of equilibrium points}\label{Determination of equilibrium points}
The Hamiltonian function, as given in \citet{2008ChJAA...8..108H}, can be expressed by equation (23).
 \begin{eqnarray}\label{eq_23}
    \text{$H$=}\frac{1}{2}\mathbf{\dot{r}}\cdot \mathbf{\dot{r}}+\text{V}\left( \mathbf{r} \right)\
    \end{eqnarray}\    
and, because \textbf{\(\omega\)} is time-invariant, we can conclude that $H$ will also be time-invariant and it is called the Jacobi constant. Redefining $H=C$, the region that prohibits particle circulation is given by $V\left( r \right)>C$, while the region that allows the particle to circulate satisfies $V\left( r \right)<C$. Lastly, the equation $V\left( r \right)=C$ implies that the particle is stationary in relation to the system fixed to the asteroid, providing zero velocity surfaces. Therefore, equilibrium points are the critical points of effective potential $\text{V}\left( \mathbf{r} \right)$, that is $\nabla \text{V}\left( x,y,z \right)\text{=}\,\text{0}$, so they have to satisfy Equation (24).
\begin{eqnarray}\label{eq_24}
   \frac{\partial \text{V}\left( x,y,z \right)}{\partial x}=\frac{\partial \text{V}\left( x,y,z 
   \right)}{\partial y}=\frac{\partial \text{V}\left( x,y,z \right)}{\partial z}=0
    \end{eqnarray}\
These algebraic equations were solved numerically using Mathematica, determining the coordinates of the equilibrium points, designated by ${{\left( {{x}_{{E}}},{{y}_{E}},{{z}_{E}} \right)}^{\text{T}}}$. It should be noted that the positions, stabilities and topological classification of the equilibrium points vary depending on the shape and angular velocity of the asteroid.\

In order to study its stability, the effective potential is expanded in a Taylor series in the neighborhood of that point, allowing the linearization of the equations of motion, as well as obtaining the associated characteristic equation, whose roots \textbf{\(\lambda\)}, also called eigenvalues, provide fundamental information about their stability. Equation (25) defines $P\left( \lambda  \right)$.
\begin{eqnarray}\label{eq_25}
P\left( \lambda  \right)=\left| \begin{matrix}
   {{\lambda }^{2}}+{{V}_{xx}} & -2\omega \lambda +{{V}_{xy}} & {{V}_{xz}}  \\
   2\omega \lambda +{{V}_{xy}} & {{\lambda }^{2}}+{{V}_{yy}} & {{V}_{yz}}  \\
   {{V}_{xz}} & {{V}_{yz}} & {{\lambda }^{2}}+{{V}_{zz}}  \\
\end{matrix} \right|=0
    \end{eqnarray}\
In Equation (25) we have
\begin{eqnarray}\label{eq_26}
  & {{\text{V}}_{xx}}={{\left( \frac{{{\partial }^{2}}\text{V}}{\partial {{x}^{2}}} \right)}_{E}}\,\,\,\,\,\,\,{{\text{V}}_{xy}}={{\left( \frac{{{\partial }^{2}}\text{V}}{\partial x\partial y} \right)}_{E}}\\  
 & {{\text{V}}_{yy}}={{\left( \frac{{{\partial }^{2}}\text{V}}{\partial {{y}^{2}}} \right)}_{E}}\,\,\,\,\,\,{{\text{V}}_{yz}}={{\left( \frac{{{\partial }^{2}}\text{V}}{\partial y\partial z} \right)}_{E}} \\ 
 & {{\text{V}}_{zz}}={{\left( \frac{{{\partial }^{2}}\text{V}}{\partial {{z}^{2}}} \right)}_{E}}\,\,\,\,\,\,{{\text{V}}_{xz}}={{\left( \frac{{{\partial }^{2}}\text{V}}{\partial x\partial z} \right)}_{E}}  
    \end{eqnarray}\
here ${{\text{V}}_{pq}}={{\left( \frac{{{\partial }^{2}}\text{V}}{\partial p\partial q} \right)}_{E}}$, $p,q=x,y,z$, is the value of the derivative calculated at the equilibrium point.

\section{Applications to irregular celestial bodies}\label{Applications to irregular celestial bodies}

    In this section, after applying the Potential Series Expansion Method (PSEM) to model the gravitational field around asteroids (87) Sylvia, (101955) Bennu, (25143) Itokawa, and (99942) Apophis, we determine the equilibrium points, analyze the corresponding stability, and use PSEM to calculate the potential on a grid of 1,002,000 points near the target. We compare our results with the classical polyhedral method described by \citet{tsoulis_2001} and the Tetrahedron Center (TC) model presented by \citet{aljbaae_2021}.

    The stability of equilibrium points, a well-documented phenomenon in celestial mechanics, is used here as a benchmark to validate the accuracy of PSEM. To enhance clarity and avoid redundancy, we summarize the common trends observed across all analyzed bodies while focusing on the unique findings specific to each asteroid.

    To carry out our dynamic studies, we initially need to refine the coordinate data of the vertices for the models of the four analyzed asteroids. We performed translations and rotations to ensure that the center of mass and the main inertia axes of each asteroid aligned with the origin and coordinate axes of a system fixed on the asteroids.

    To demonstrate the efficiency of our method, we calculated the gravitational potential near our targets. We conducted a series of tests comparing the potential $U_{PSEM}$ calculated by PSEM at different orders (e.g., 5, 6, 7,...) and the gravitational model presented by \citet{aljbaae_2021}, $U_{TC20}$, which divides the asteroid into 20 layers of equal density, with the classical polyhedral method $U_{T}$ calculated by \citet{tsoulis_2001}. We compute the relative errors between $U_{PSEM}$ (at various orders) or $U_{TC20}$ and
    $U_{T}$ as follows, shown in Equation (29).
    \begin{eqnarray}\label{eq_re}
        RE = \frac{U-U_{T}}{U_{T}}
    \end{eqnarray}

    where $RE$ is the relative error and $U$ represents either $U_{PSEM}$ or $U_{TC20}$. Across all analyzed bodies, the equilibrium points and their stability, as determined using PSEM, align well with established results in the literature. This demonstrates the reliability and accuracy of the method in reproducing gravitational field features. Additionally, the PSEM achieves significant computational efficiency, considerably reducing the execution time compared to the classical polyhedral approach. These results validate the versatility of the PSEM for modeling gravitational fields of both large and small celestial bodies with irregular shapes.

\subsection{Study of the asteroid (87) Sylvia}\label{Study of the asteroid (87) Sylvia}

    To study asteroid (87) Sylvia, we used a non-convex polyhedral shape model with 800 triangular faces and 402 vertices. The model is available in the Inversion Techniques Asteroid Model Database (DAMIT\footnote{\href{http://astro.troja.mff.cuni.cz/projects/damit}{http://astro.troja.mff.cuni.cz/projects/damit}}), as referenced by \citet{durech_2010, marchis_2006, hanus_2013}. The shape of the asteroid, in kilometers, spans the following ranges in the primary directions: $(-192.6082,176.1018)\times(-128.5673,127.2156)\times(-121.4823,124.4925)$. To compute the integrals, we used the method proposed by \citet{Lien_1984}.

    Assuming a homogeneous structure with a uniform density of 1.373 g/cm$^{3}$ \citep{berthier_2014} and a volume of $1.07$x$10^7$ km$^3$, we found that the mass is $1.47$x$10^{19}$ kg. We adopted a rotation period of 5.184 hours \citep{marchis_2006}.

\begin{figure}[ht]   
        \begin{center}
            \includegraphics[width=0.6\linewidth]{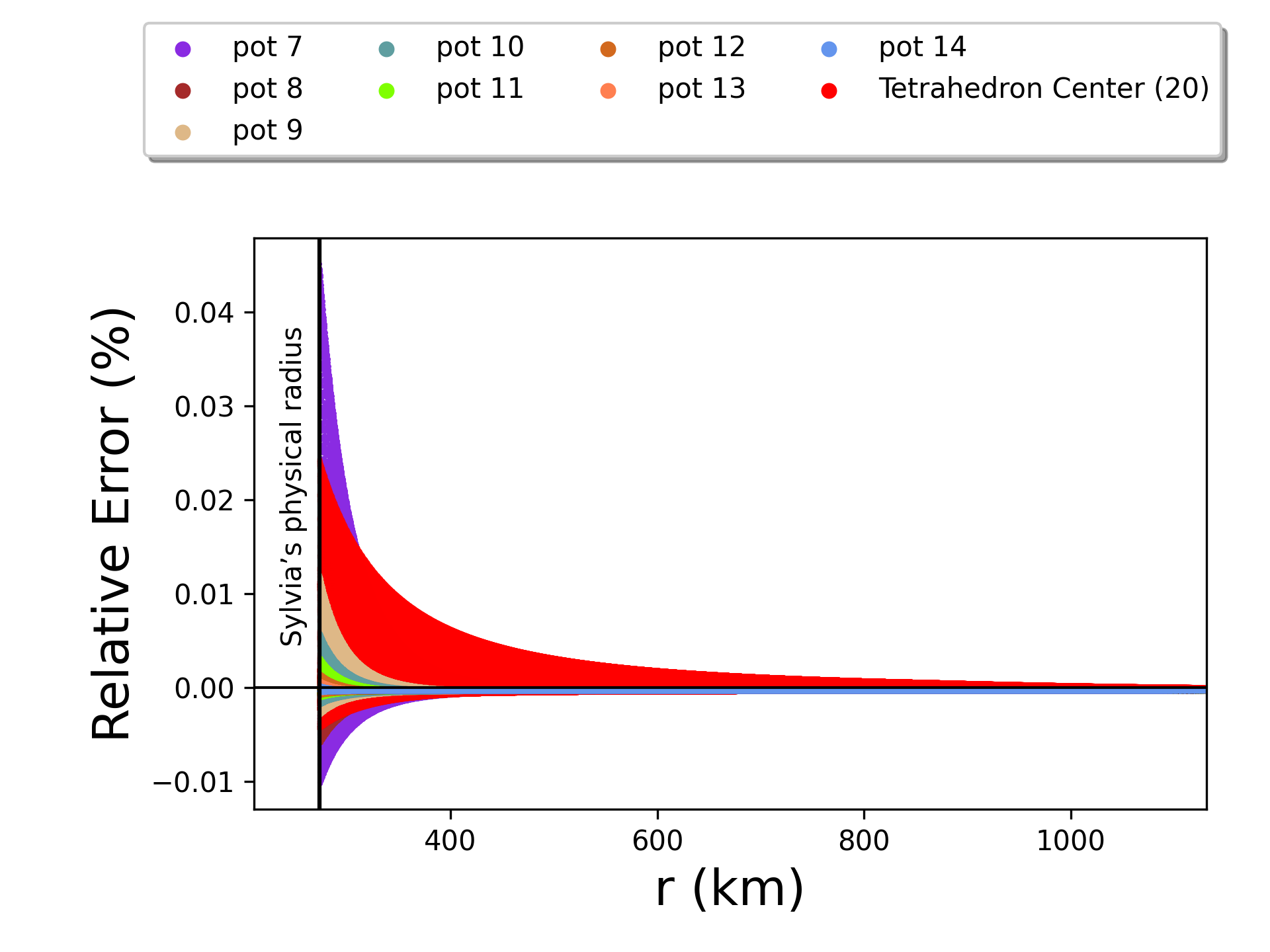}
        \end{center}
        \caption{Relative error of the gravitational potential $U_{PSEM}$ or $U_{TC20}$ with respect to the classical polyhedral method ($U_{T}$)} \label{pot_relativ_error}
    \end{figure}

    Our results are presented in Figure \ref{pot_relativ_error}, where $RE$ is the relative error, and $U$ represents either $U_{PSEM}$ or $U_{TC20}$.
    We observe good agreement with these models outside the body (to the right of the red line). The results indicate that our model provides better accuracy than the approach in \citet{aljbaae_2021}, provided the potential is developed to an order higher than 9. Table \ref{table_execution_time} shows the CPU time required to calculate the potential of a grid of 1,002,000 points outside the asteroid using a Pentium 3.10 GHz CPU. Our method significantly reduces computational processing time compared to the classical polyhedral method while maintaining a high level of accuracy.     
    \begin{table}[!htp]
    \caption{Execution time for calculating the gravitational potential on a grid of 1,002,000 points close to Sylvia using a Pentium 3.60 GHz CPU.} \label{table_execution_time}
            \begin{tabular}{lll}
                \hline
                \citet{tsoulis_2001} &  \citet{aljbaae_2021} & This work (PSEM 14) \\
                \hline
                31m34.526s &  1m52.598s & 0m16.979s \\
                \hline
                   Times faster compared to Tsoulis        &  16.8256         &   111.58054       \\
                \hline
                   Position vector accuracy EP   &     0.094       &      0.073    \\
                \hline  
            \end{tabular}
    \end{table}

\begin{figure}[ht]
        \begin{center}
            \includegraphics[width=0.6\linewidth]{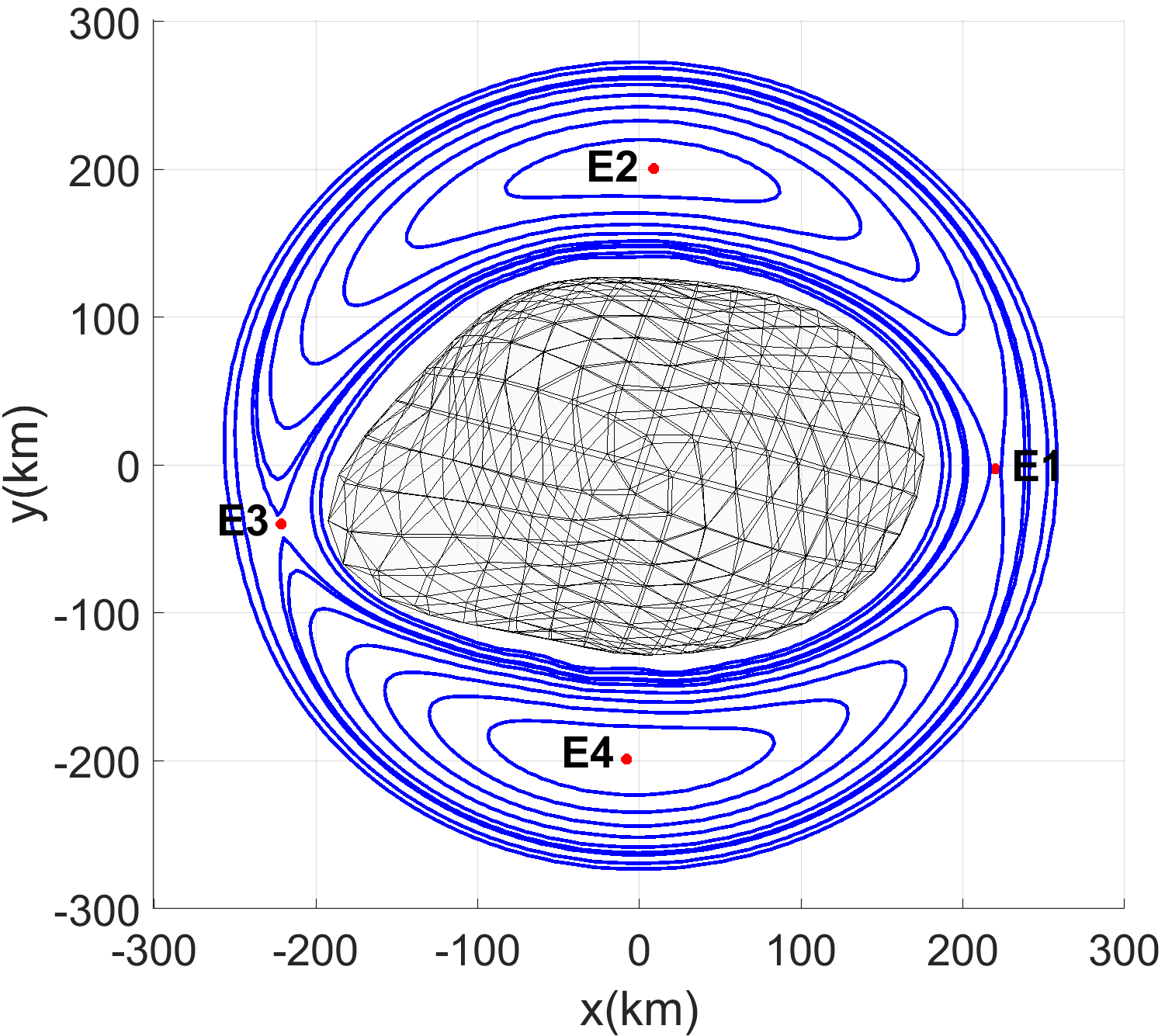}
        \end{center}
         
        \caption{Zero-velocity curves and equilibrium points of (87) Sylvia} \label{fig_eq_points}
    \end{figure}

    We used our method to find the zero-velocity surfaces and equilibrium points of the target and assess their stability. For more information on this energy equation, refer to \citet{aljbaae_2021}, \citet{Jiang_2014} and \citet{Wang_2014}. Figure \ref{fig_eq_points} shows the projection of the zero-velocity surfaces onto the xy-plane. The position and relative error of each equilibrium point's vector position relative to the classical polyhedral method (Tsoulis, 2001) are provided in Table \ref{equlibrium_points_Sylvia}. Pot11 and Pot12 refer to the potentials expanded to orders 11 and 12, respectively, per Equation (20).

    \begin{table}[!htp]
        \caption{Position of the equilibrium points (in km) outside (87) Sylvia, using the potential series expansion method, and their relative errors ($RE$) compared to the classical polyhedral method.} \label{equlibrium_points_Sylvia}
            \begin{tabular}{|ll|l|l||l|l|}
                \hline
                  & & Pot 11 & $|RE| (\%)$ &  Pot 12 & $|RE| (\%)$\\
                \hline
                \multirow{3}{*}{}& $x$ &  220.191  &       & 220.197 &\\
                $E_1$            & $y$ & -3.961    & 0.013 & -3.299 & 0.011\\
                                 & $z$ & -1.005    &       & -1.037  &  \\
                
                \hline
                \multirow{3}{*}{}& $x$ & 6.837    &      & 6.188   &\\
                $E_2$            & $y$ & 200.652  & 0.043 & 200.598 & 0.0054\\
                                 & $z$ & 0.227    &      & 0.2307 & \\

                                 \hline
                \multirow{3}{*}{}& $x$ & -221.705 &      & -221.755 &\\
                $E_3$            & $y$ & -41.382  & 0.116 & -41.648 & 0.073\\
                                 & $z$ & -0.128    &      & -0.141  &\\
                \hline
                \multirow{3}{*}{}& $x$ & -9.448  &      & -8.681   &\\
                $E_4$            & $y$ & -199.343 & 0.072 & -199.301 & 0.034\\
                                 & $z$ & -0.996   &      & -1.012  & \\

                \hline
            \end{tabular}
    \end{table}

    \begin{table}[!htp]
        \caption{Eigenvalues of the equilibrium points around asteroid (87) Sylvia.} \label{Eigenvalues_Sylvia}
            \begin{tabular}{|l|l|l|l|l|}
                \hline
                  $\times 10^{-4}$& $E_1$ & $E_2$ &  $E_3$ & $E_4$\\
                \hline
                $\lambda_1$ & 2.536 & 3.413$i$ & 3.376 & 3.456$i$\\
                \hline
                $\lambda_2$  & -2.536 & -3.413$i$ & -3.376 & -3.456$i$\\
                \hline
                $\lambda_3$  & 3.903$i$ & -0.965 + 2.538$i$ & 4.201$i$ & -1.277 + 2.645$i$ \\
                \hline
                $\lambda_4$  & -3.903$i$ & -0.965 - 2.538$i$ & -4.201$i$ & -1.277 - 2.645$i$\\
                \hline
                $\lambda_5$  & 3.725$i$ & 0.965 + 2.538$i$ & 4.052$i$ & 1.277 + 2.645$i$\\
                \hline
                $\lambda_6$  & -3.725$i$ & 0.965 - 2.538$i$ & -4.052$i$ & 1.277 - 2.645$i$\\                \hline
            \end{tabular}
    \end{table}

    According to the topological classification of \citet{Jiang_2014} and \citet{Wang_2014}, Table \ref{Eigenvalues_Sylvia} shows that points $E_1$ and $E_3$ have two real eigenvalues and four pure imaginary eigenvalues, classifying them as Case 2: saddle-center-center. This reveals two families of periodic orbits and a family of quasi-periodic orbits in their vicinity. Points $E_2$ and $E_4$, however, present two pure imaginary eigenvalues and four complex eigenvalues, falling under Case 5: sink-source-center, indicating the presence of only a family of periodic orbits near each point.

\subsection{Study of the asteroid (101955) Bennu}\label{Study of the asteroid (101955) Bennu.}\

    To study asteroid (101955) Bennu, we adopted a non-convex polyhedral shape model with 12,288 triangular faces, available in the Planetary Data System (PDS\footnote{\href{https://pds.nasa.gov/}{https://pds.nasa.gov/}}, \citep{2013PDSS..211.....N}). The dimensions of this asteroid, in kilometers, are  $(-0.2800, 0.2787)\times(-0.2617, 0.2671)\times(-0.2446, 0.2540)$. To calculate the integrals, we used the method by \citet{Lien_1984}. Assuming a homogeneous structure with a uniform density of 1.25 g/cm$^{3}$ \citep{2014Icar..235....5C} and a volume of 0.0623 km${^3}$. The mass is calculated as $7.79$x$10^{10}$ kg. We adopted a rotation period of 4.297 hours \citep{2013PDSS..211.....N}.

    \begin{figure}[ht]
        \begin{center}
            \includegraphics[width=0.6\linewidth]{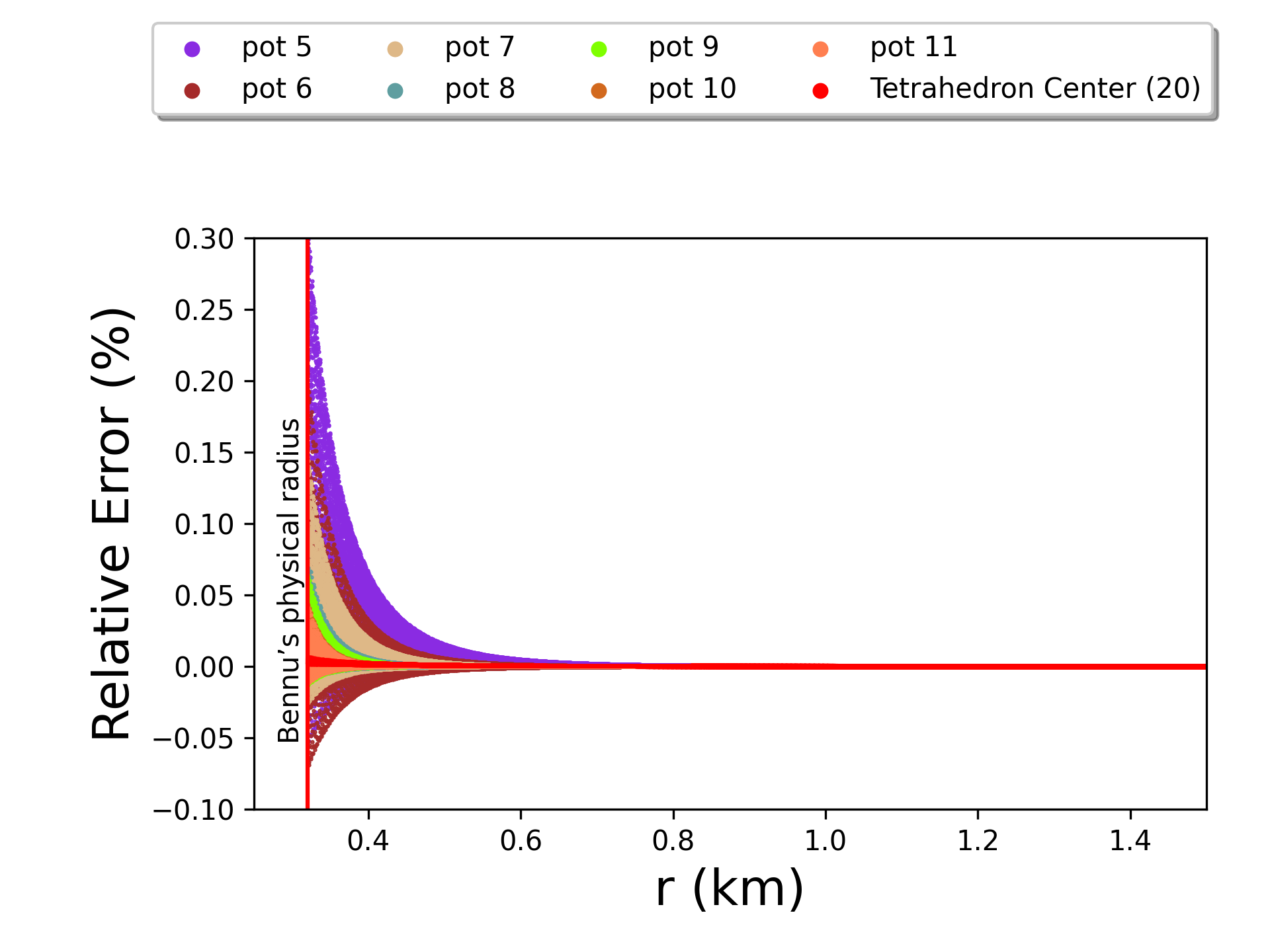}
        \end{center}
         
        \caption{Relative error of the gravitational potential $U_{PSEM}$ or $U_{TC20}$ compared to the classical polyhedral method ($U_{T}$)} \label{pot_relativ_error_BENNU}
    \end{figure}  

     \begin{table}[!htp]
    \caption{Execution time for calculating the gravitational potential on a grid of 1,002,000 points close to Bennu using a Pentium 3.60 GHz CPU.} \label{table_execution_time_BENNU}
            \begin{tabular}{lll}
                \hline
                \citet{tsoulis_2001} &  \citet{aljbaae_2021} & This work (PSEM 11) \\
                \hline
                42m2.000s &  0m30.180s & 0m3.190s \\
                \hline
                  Times faster compared to Tsoulis      &    83.5653        &    790.5956      \\
                \hline
                   Position vector accuracy EP   &     0.062       &     0.12     \\
                \hline
            \end{tabular}
    \end{table}
    
     \noindent Our results are presented in Figure \ref{pot_relativ_error_BENNU}, where $RE$ is the relative error, and $U$ is either $U_{PSEM}$ or $U_{TC20}$. We observe good agreement between these models outside the body (right side of the red line). Our model yields better results than the approach presented by \citet{aljbaae_2021} if the potential is expanded to an order higher than 9. Table \ref{table_execution_time_BENNU} shows the CPU time required to compute the potential of a grid of 1,002,000 points outside the asteroid using a Pentium 3.60 GHz CPU. Notably, our method significantly reduces computation processing time compared to the classical polyhedral method while maintaining a high level of accuracy.

\begin{figure}[ht]
        \begin{center}
            \includegraphics[width=0.6\linewidth]{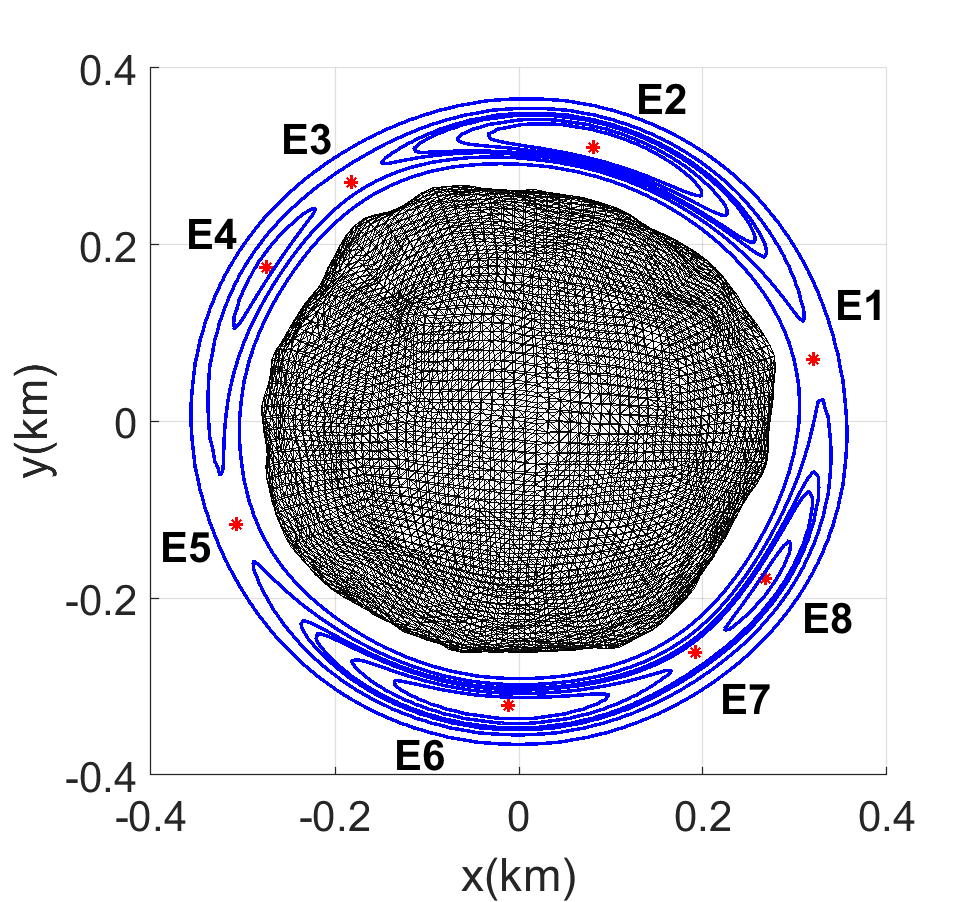}
        \end{center}
        \caption{Zero-velocity curves and equilibrium points of (101955) Bennu.} \label{fig_eq_points_BENNU}
    \end{figure}

     We applied our method to find the zero-velocity surfaces and equilibrium points of the target, assessing their stability. For more details on this energy equation, refer to \citet{aljbaae_2021}, \citet{Jiang_2014} and \citet{Wang_2014}. Figure \ref{fig_eq_points_BENNU} shows the projection of the zero-velocity surfaces onto the xy plane. The position and relative error of each equilibrium point's vector position compared to the classical polyhedral method \citep{tsoulis_2001} are provided in Table \ref{equlibreum_points_Bennu}. Pot10 and Pot11 refer to the potentials expanded to orders 10 and 11, respectively, per Equation (20).

  \begin{table}[!htp]
        \caption{Position of the equilibrium points (in km) outside (101955) Bennu using the Potential Series Expansion Method and their relative errors ($RE$) compared to the classical polyhedral method.} \label{equlibreum_points_Bennu}
            \begin{tabular}{|ll|l|l||l|l|}
            \hline
                & & Pot 10 & $|RE| (\%)$ &  Pot 11 & $|RE| (\%)$\\
                \hline
                \multirow{3}{*}{}& $x$ & 0.320084  &         & 0.320299&\\
                $E_1$            & $y$ & 0.071455  & 0.13214 & 0.070627 & 0.12291\\
                                 & $z$ & -0.00061  &         &-0.00064  &  \\
                
                \hline
                \multirow{3}{*}{}& $x$ & 0.080495    &      &0.081055   &\\
                $E_2$            & $y$ & 0.310212  &0.00653 &0.310079 & 0.00222\\
                                 & $z$ & 0.001453    &      & 0.001456 & \\

                \hline
                \multirow{3}{*}{}& $x$ & -0.181241 &      & -0.182605&\\
                $E_3$            & $y$ & 0.271724  & 0.09276 & 0.270832 & 0.08712\\
                                 & $z$ & -0.000054    &      & -0.000066  &\\
                \hline
                \multirow{3}{*}{}& $x$ & -0.272636  &      &-0.273951   &\\
                $E_4$            & $y$ & 0.175824 &0.04120 &0.173731 &0.03518\\
                                 & $z$ & -0.001907  &      & -0.001939  & \\

                                 \hline
                \multirow{3}{*}{}& $x$ & -0.307399  &      &-0.306813   &\\
                $E_5$            & $y$ & -0.114678 &0.04371 &-0.116349 &0.03169\\
                                 & $z$ & -0.002318   &      & -0.002425  & \\

                                 \hline
                \multirow{3}{*}{}& $x$ & -0.010622  &      &-0.011002   &\\
                $E_6$            & $y$ & -0.321361 &0.03645 &-0.321341 &0.03395\\
                                 & $z$ & -0.002497   &      & -0.002405  & \\

                                 \hline
                \multirow{3}{*}{}& $x$ & 0.193843  &      &0.192142   &\\
                $E_7$            & $y$ & -0.260166 &0.03380 &-0.261382 &0.03069\\
                                 & $z$ & -0.001056   &      & -0.001163  & \\

                                 \hline
                \multirow{3}{*}{}& $x$ & 0.269660  &      &0.268367   &\\
                $E_8$            & $y$ & -0.176038 &0.02938 &-0.178066 &0.02727\\
                                 & $z$ & -0.0009202   &      & -0.0009307  & \\

                \hline
            \end{tabular}
    \end{table}

 \begin{table}[!htp]
        \caption{Eigenvalues of the equilibrium points around the asteroid (101955) Bennu.} \label{Eigenvalues_Bennu}
            \begin{tabular}{|l|l|l|l|l|}
                \hline
                  $\times 10^{-4}$& $E_1$ & $E_2$ &  $E_3$ & $E_4$\\
                \hline
                $\lambda_1$ & 2.168 & 4.436$i$ & 1.999 & 4.569$i$\\
                \hline
                $\lambda_2$  & -2.168 & -4.436$i$ & -1.999 & -4.569$i$ \\
                \hline
                $\lambda_3$  & 4.082$i$ & 0.713 + 2.676$i$ & 3.975i & 0.755 + 2.575$i$\\
                \hline
                $\lambda_4$  & -4.082$i$ & 0.713 - 2.676$i$ & -3.975$i$ & 0.755 - 2.575$i$\\
                \hline
                $\lambda_5$  & 4.585$i$ & -0.713 + 2.676$i$ & 4.602$i$ & -0.755 + 2.575$i$\\
                \hline
                $\lambda_6$  & -4.585$i$ & -0.713 - 2.676$i$ & -4.602$i$ & -0.755 - 2.575$i$\\                \hline
                & $E_5$ & $E_6$ &  $E_7$ & $E_8$\\
                \hline
               $\lambda_1$ & 1.937  & 2.012$i$  &  1.830  & 4.416$i$ \\
                \hline
               $\lambda_2$ & -1.937  & -2.012$i$  &  -1.830  & -4.416$i$ \\
                \hline
               $\lambda_3$ &  3.889$i$ & 3.033$i$  &  4.018$i$  & 0.695 + 2.688$i$ \\
                \hline
               $\lambda_4$ & -3.889$i$  & -3.033$i$  &  -4.018$i$  & 0.695 - 2.688$i$ \\
                \hline
               $\lambda_5$ &  4.650$i$ & 4.443$i$  & 4.494$i$   & -0.695 + 2.688$i$ \\
                \hline
               $\lambda_6$ & -4.650$i$  &  -4.443$i$ &  -4.494$i$  & -0.695 - 2.688$i$ \\               \hline
            \end{tabular}
    \end{table}

    Examining Table \ref{Eigenvalues_Bennu}, and following the topological classification established by \citet{Jiang_2014} and \citet{Wang_2014}, we found that points $E_1$, $E_3$, $E_5$ and $E_7$ have two real eigenvalues and four pure imaginary eigenvalues, placing them in Case 2, saddle-center-center. This configuration reveals two families of periodic orbits and a family of quasi-periodic orbits near each of these points. On the other hand, points $E_2$, $E_4$ and $E_8$ have two pure imaginary eigenvalues and four complex eigenvalues, classifying them in Case 5, sink-source-center, which suggests the presence of only one family of periodic orbits in the vicinity of each point. Finally, the linearly stable equilibrium point $E_6$ presents three pairs of purely imaginary eigenvalues, indicating three families of periodic orbits and placing it in Case 1.

 \subsection{Study of the asteroid (25143) Itokawa}\label{Study of the asteroid (25143) Itokawa.}\

    To study asteroid (25143) Itokawa, we adopted a non-convex polyhedral shape model with 12,192 triangular faces. This model is available in the Planetary Data System (PDS\footnote{\href{https://pds.nasa.gov/}{https://pds.nasa.gov/}}, \citep{2008PDSS...92.....G}). The total dimensions of the asteroid's shape in kilometers are $(-0.2597, 0.3049)\times(-0.1581, 0.1418)\times(-0.1199, 0.1239)$. As with the previous asteroids, we used the method of \citet{Lien_1984} to calculate the integrals relating to the physical properties of Itokawa. Assuming a homogeneous structure with a uniform density of 1.98 g/cm${^3}$ \citep{2006Sci...312.1344A, Scheeres2006TheAD} and a volume of 0.0178 km${^3}$, the mass is calculated to be $7.79$x$10^{10}$ kg. We adopted a rotation period of 12.132 hours \citep{2006Sci...312.1344A, Scheeres2006TheAD}.

\begin{figure}[ht]
        \begin{center}
            \includegraphics[width=0.6\linewidth]{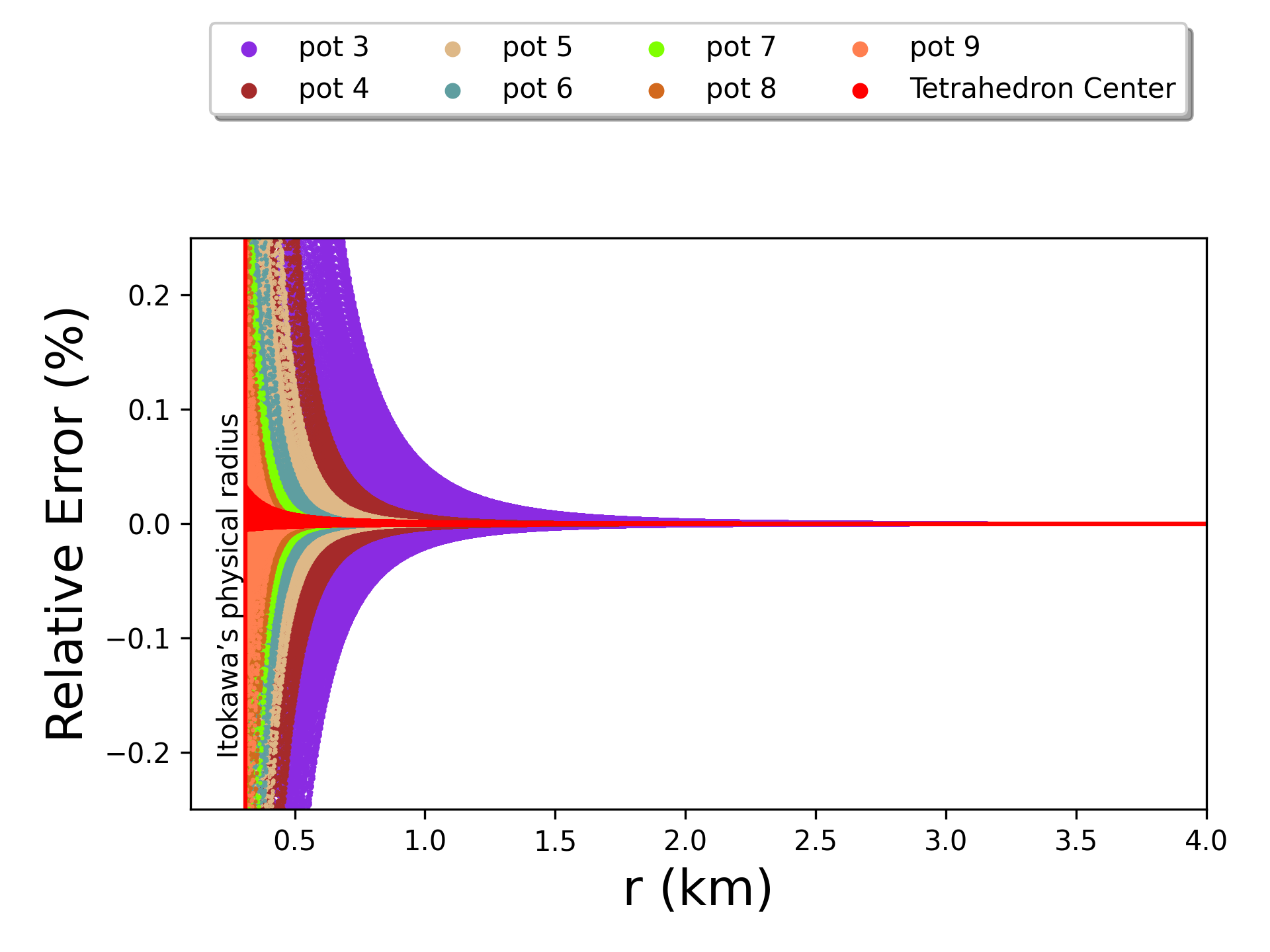}
        \end{center}

        \caption{Relative error of the gravitational potential $U_{PSEM}$ or $U_{TC20}$ with respect to the classical polyhedral method ($U_{T}$)} \label{pot_relativ_error_Itokawa}
    \end{figure}

\begin{figure}[ht]
        \begin{center}
            \includegraphics[width=0.6\linewidth]{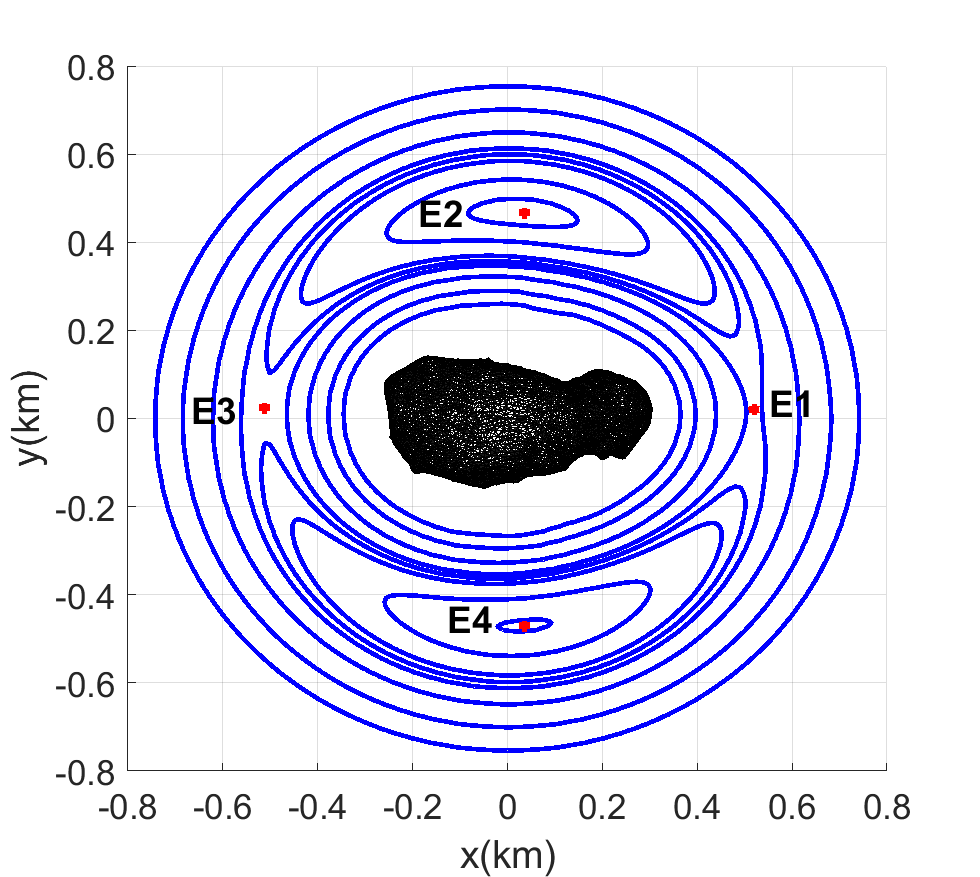}
        \end{center}
        \caption{Zero-velocity curves and equilibrium points of (25143) Itokawa.} \label{fig_eq_points_Itokawa}
        \end{figure}

    Our results are presented in Figure \ref{pot_relativ_error_Itokawa}, where $RE$ is the relative error, and  $U$ is either $U_{PSEM}$ or $U_{TC20}$. We observe good agreement between these models outside the body (to the right of the red line). Our model provides better results than the approach presented in \citet{aljbaae_2021} if the potential is developed with an order higher than 9.
    In Table \ref{table_execution_time_Itokawa}, we present the CPU time required to compute the potential of a grid of 1,002,000 points outside the asteroid using a Pentium 3.10 GHz CPU. Our method significantly reduces computational processing time compared to the classical polyhedral method, while maintaining a high level of accuracy.
    \begin{table}[!htp]
    \caption{Execution time for calculating the gravitational potential on a grid of 1,002,000 points close to (25143) Itokawa using a Pentium 3.60 GHz CPU} \label{table_execution_time_Itokawa}
            \begin{tabular}{lll}
                \hline
                \citet{tsoulis_2001} &  \citet{aljbaae_2021} & This work (PSEM 9) \\
                \hline
                475m36s &  27m29s & 0m32s \\
                \hline
                   Times faster compared to Tsoulis    &    17.3050        &   891.7500       \\
                \hline
                   Position vector accuracy EP   &    0.038        &      0.010    \\
                \hline
            \end{tabular}
    \end{table}
    We used our method to obtain the zero-velocity surfaces and equilibrium points of the target, then assessed their stability. For more details on the energy equation, refer to \citet{aljbaae_2021}, \citet{Jiang_2014} and \citet{Wang_2014}. In Figure \ref{fig_eq_points_Itokawa}, we present the projection of the zero-velocity surfaces onto the xy-plane. The position and relative error of each equilibrium point's vector position, compared to the classical polyhedral method \citep{tsoulis_2001}, are given in Table \ref{equlibreum_points_Itokawa}. Pot 8 and Pot 9 refer to the potentials developed up to orders 8 and 9, respectively, per Equation 19.
    \begin{table}[!htp]
        \caption{Position of the equilibrium points (in km) outside (25143) Itokawa using the Potential Series Expansion Method and their relative errors ($RE$) compared to the classical polyhedral method.} \label{equlibreum_points_Itokawa}
            \begin{tabular}{|ll|l|l||l|l|}
                \hline
                  & & Pot 8 & $|RE| (\%)$ &  Pot 9 & $|RE| (\%)$\\
                \hline
                \multirow{3}{*}{}& $x$ &  0.520675  &       & 0.520710&\\
                $E_1$            & $y$ & 0.019632   & 0.0081720 & 0.019729 & 0.0005064\\
                                 & $z$ & -0.008670    &       & -0.008796  &  \\

                \hline
                \multirow{3}{*}{}& $x$ & 0.035558    &      & 0.035332   &\\
                $E_2$            & $y$ & 0.467753  & 0.0194685 & 0.467857 & 0.0098828\\
                                 & $z$ & 0.001936    &      & 0.001939 & \\

                                 \hline
                \multirow{3}{*}{}& $x$ & -0.511872 &      & -0.511832&\\
                $E_3$            & $y$ & 0.0248312  & 0.0103370 & 0.025436 & 0.0084465\\
                                 & $z$ & -0.0043722    &      & -0.004505  &\\
                \hline
                \multirow{3}{*}{}& $x$ & 0.0378808  &      & 0.036503   &\\
                $E_4$            & $y$ & -0.4705003 & 0.0076195 & -0.470619 & 0.0056761\\
                                 & $z$ & 0.0015133   &      & 0.001469  & \\

                \hline
            \end{tabular}
    \end{table}
   \begin{table}[!htp]
        \caption{Eigenvalues of the equilibrium points around asteroid (25143) Itokawa.} \label{Eigenvalues_Itokawa}
            \begin{tabular}{|l|l|l|l|l|}
                \hline
                  $\times 10^{-4}$& $E_1$ & $E_2$ &  $E_3$ & $E_4$\\
                \hline
                $\lambda_1$ & 1.165 & 1.446$i$ & 0.928 & 1.453$i$\\
                \hline
                $\lambda_2$  & -1.165 & -1.446$i$ & -0.928 & -1.453$i$\\
                \hline
                $\lambda_3$  & 1.606$i$ & -0.497 + 1.127$i$ & 1.565$i$ & -0.418  + 1.091$i$ \\
                \hline
                $\lambda_4$  & -1.606$i$ & -0.497 - 1.127$i$ & -1.565$i$ & -0.418 - 1.091$i$\\
                \hline
                $\lambda_5$  & 1.708$i$ & 0.497 + 1.127$i$ & 1.598$i$ & 0.418  + 1.091$i$\\
                \hline
                $\lambda_6$  & -1.708$i$ & 0.497 - 1.127$i$ & -1.598$i$ & 0.418 - 1.091$i$\\                \hline
            \end{tabular}
    \end{table}

    Examining Table \ref{Eigenvalues_Itokawa}, and following the topological classification of \citet{Jiang_2014} and \citet{Wang_2014}, we found that points $E_1$ and $E_3$ have two real eigenvalues and four pure imaginary eigenvalues, placing them in Case 2, saddle-center-center. This configuration reveals two families of periodic orbits and a family of quasi-periodic orbits near each of these points. Points $E_2$ and $E_4$, on the other hand, have two pure imaginary eigenvalues and four complex eigenvalues, placing them in Case 5, sink-source-center, which indicates the presence of only one family of periodic orbits in the neighborhood of each point.

\subsection{Study of the asteroid (99942) Apophis}\label{Study of the asteroid (99942) Apophis.}

    For the study of asteroid Apophis, we based our analysis on the model given by \citep{2014Icar..233...48P}, available on the 3D Asteroid Catalog website and later refined by \citep{2018Icar..300..115B}. We used a polyhedral model containing 2,000 vertices and 3,996 faces (accessible at 3d-asteroids.space\footnote{\href{https://3d-asteroids.space/asteroids/99942-Apophis}{https://3d-asteroids.space/asteroids/99942-Apophis}}), referenced in \citep{2014Icar..233...48P, 2018Icar..300..115B, 2021RoAJ...31..241A}. The asteroid's total shape dimensions in the primary directions, in kilometers, are $(-0.2586,0.2136)\times(-0.1983,0.2039)\times(-0.1689,0.1980)$. We used the method developed by \citet{Lien_1984} to calculate the integrals. Assuming a homogeneous structure with a uniform density of 1.75 g/cm$^{3}$ \citep{2014A&A...562A..48L}, and a volume of 0.03034 km${^3}$, we calculated the mass to be $5.31$x$10^{10}$ kg. The rotation period is adopted as 30.40 hours \citep{2014Icar..233...48P}.\\

\begin{figure}[ht]
        \begin{center}
            \includegraphics[width=0.6\linewidth]{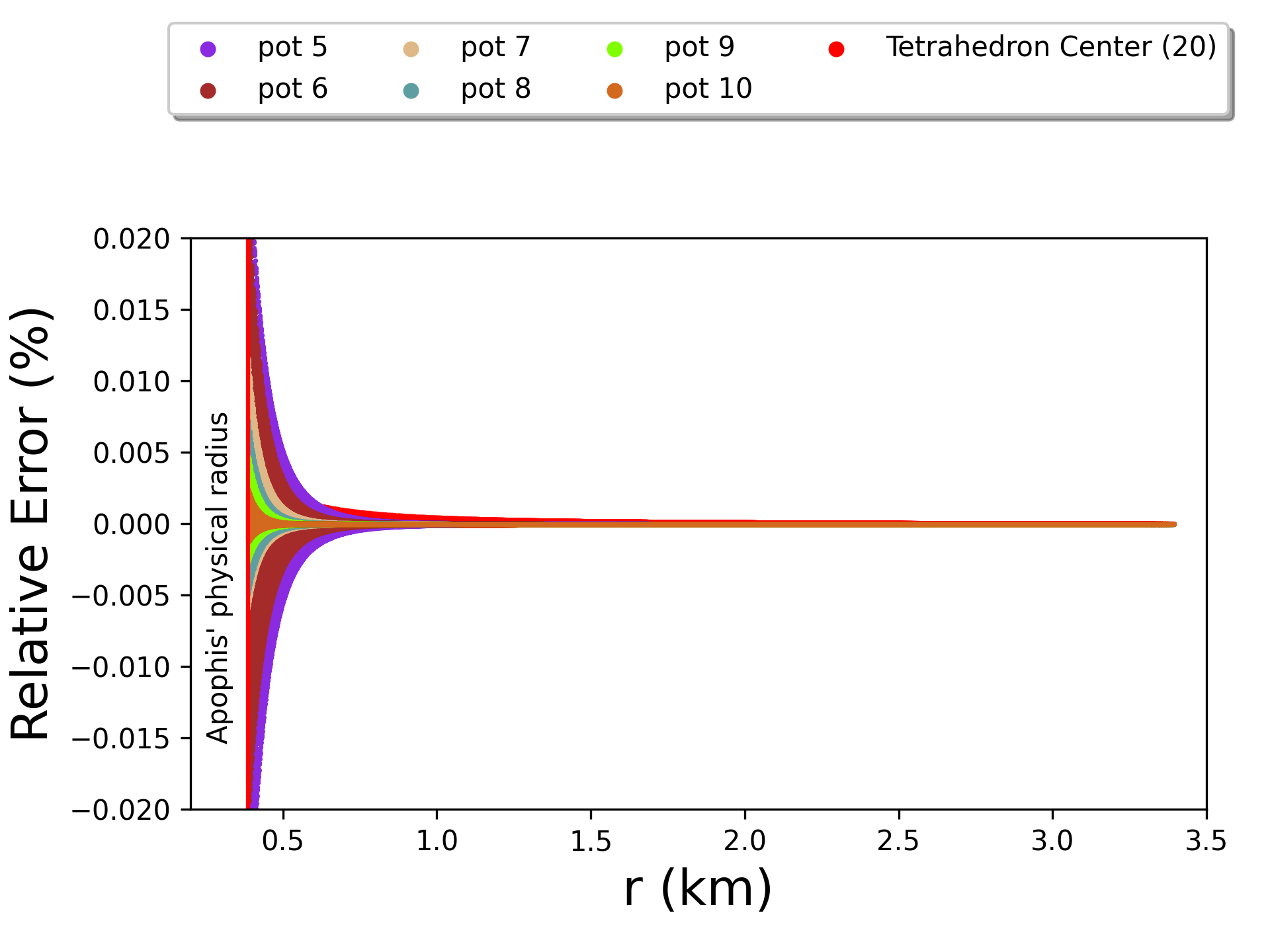}
        \end{center}
        \caption{Relative error of the gravitational potential $U_{PSEM}$ or $U_{TC20}$ compared to the classical polyhedral method ($U_{T}$)} \label{pot_relativ_error_Apophis}
    \end{figure}

    Our results are presented in Figure \ref{pot_relativ_error_Apophis}, where $RE$ is the relative error, and $U$ is either $U_{PSEM}$ or $U_{TC20}$. We observe good agreement between these models outside the body (to the right of the red line). Our model provides better results than the approach presented in \citet{aljbaae_2021} if the potential is developed to an order higher than 9. For more details about the PSEM method, refer to Appendix \ref{appendice_b}.\\

\begin{figure}[ht]   
        \begin{center}
            \includegraphics[width=0.6\linewidth]{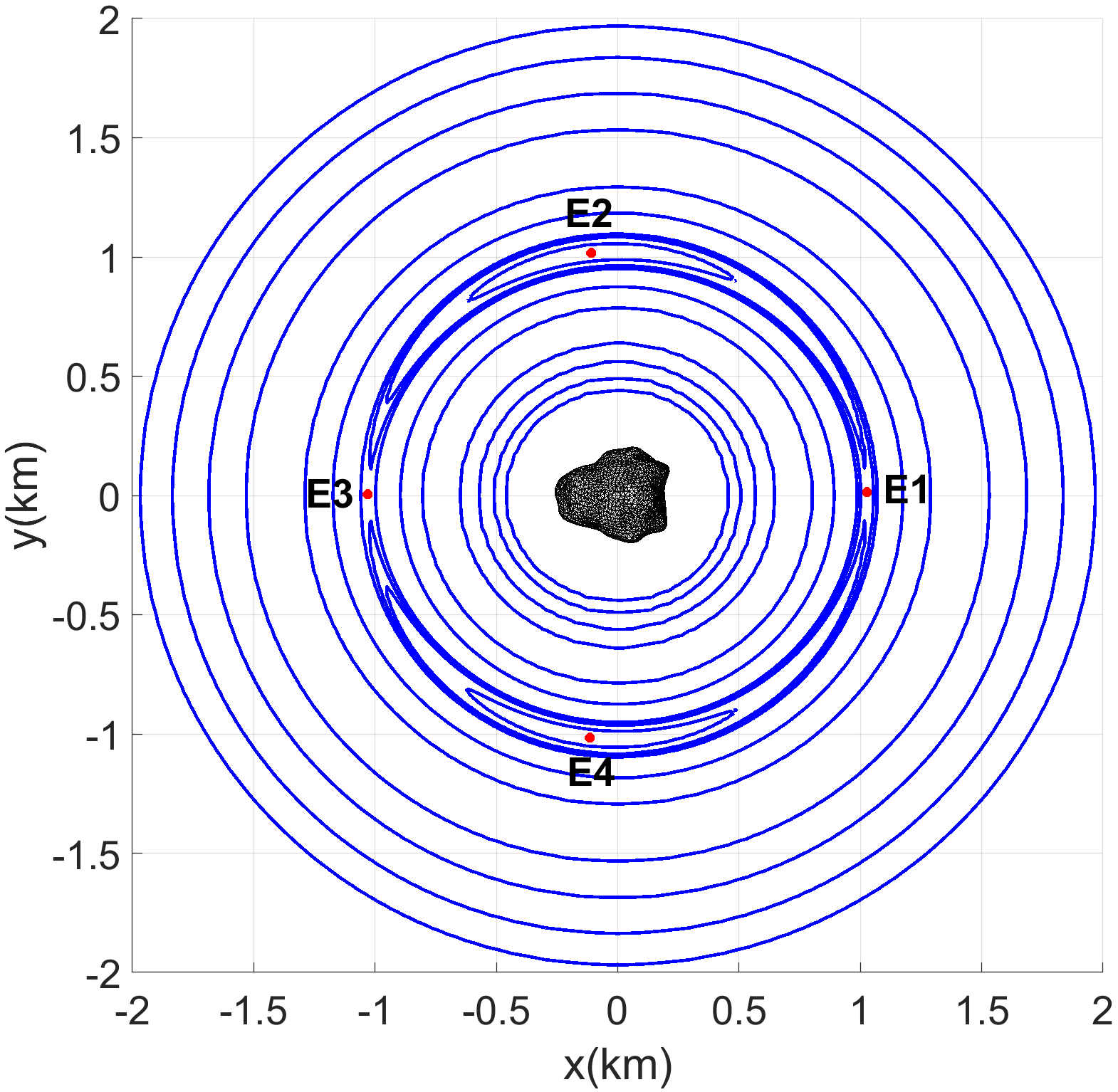}
        \end{center}
        \caption{Zero-velocity curves and equilibrium points of (99942) Apophis.} \label{fig_eq_points_Apophis}
    \end{figure}

    In Table \ref{table_execution_time_Apophis}, we present the CPU time required to compute the potential of a grid of 1,002,000 points outside the asteroid using a Pentium 3.10 GHz CPU. Notably, our method significantly reduces computational processing time compared to the classical polyhedral method while maintaining a high level of accuracy.
    \begin{table}[!htp]

        \caption{Execution time for calculating the gravitational potential on a grid of 1,002,000 points close to Apophis using a Pentium 3.60 GHz CPU.} \label{table_execution_time_Apophis}
        \begin{tabular}{lll}
            \hline
            \citet{tsoulis_2001} &  \citet{aljbaae_2021} & This work (PSEM 10) \\
            \hline
            166m54.5s &  7m43.78s & 0m15.55s \\
            \hline
               Times faster compared to Tsoulis     &    21.5932      &    644.0193      \\
            \hline
            Position vector accuracy EP   &     0.0077       &     0.0076     \\
            \hline  
        \end{tabular}
    \end{table}
    Equilibrium points play a vital role in understanding the dynamics of celestial bodies. Using the PSEM, we determined the coordinates of equilibrium points and examined their stability. Figure \ref{fig_eq_points_Apophis} presents the positions of the equilibrium points on the xy-plane. The positions of each equilibrium point's vector, along with their relative errors compared to the classical polyhedral method, are provided in Table \ref{equlibreum_points_Apophis}.

     \begin{table}[!htp]
        \caption{Position of the equilibrium points (in km) outside (99942) Apophis using the Potential Series Expansion Method and their relative errors ($RE$) compared to the classical polyhedral method.} \label{equlibreum_points_Apophis}
            \begin{tabular}{|ll|l|l|}
            \hline
                & & Pot 10 & $|RE| (\%)$ \\
                \hline
                \multirow{3}{*}{}& $x$ & 1.027942  &          \\
                $E_1$            & $y$ & -0.013557 & 0.007502 \\
                                 & $z$ & -0.00129  &          \\
                
                \hline
                \multirow{3}{*}{}& $x$ & -0.108401 &          \\
                $E_2$            & $y$ & 1.017142  & 0.007573 \\
                                 & $z$ & 0.000459  &          \\

                \hline
                \multirow{3}{*}{}& $x$ & -1.029467 &      \\
                $E_3$            & $y$ & 0.004607  & 0.007470 \\
                                 & $z$ &-0.001387  &      \\
                \hline
                \multirow{3}{*}{}& $x$ & -0.113899  &      \\
                $E_4$            & $y$ & -1.016513 & 0.007574 \\
                                 & $z$ & 0.000372  &      \\

                \hline
            \end{tabular}
    \end{table}
    
    \begin{table}[!htp]
        \caption{Eigenvalues of the equilibrium points around the asteroid (99942) Apophis.} \label{Eigenvalues_Apophis}
            \begin{tabular}{|l|l|l|l|l|}
                \hline
                  $\times 10^{-5}$& $E_1$ & $E_2$ &  $E_3$ & $E_4$\\
                \hline
                $\lambda_1$ & -0.830 & -1.123$i$ & -1.224 & -1.116$i$\\
                \hline
                $\lambda_2$  & 0.830 & 1.123$i$ & 1.224 & 1.116$i$\\
                \hline
                $\lambda_3$  & -5.762$i$ & -5.622$i$ & -5.785$i$ & -5.624$i$ \\
                \hline
                $\lambda_4$  & 5.762$i$ & 5.622$i$ & 5.785$i$ & 5.624$i$\\
                \hline
                $\lambda_5$  & -5.779$i$ & -5.749$i$ & -5.827$i$ & -5.748$i$\\
                \hline
                $\lambda_6$  & 5.779$i$ & 5.749$i$ & 5.827$i$ & 5.748$i$\\               \hline
            \end{tabular}
    \end{table}

    In addition to determining the coordinates of the equilibrium points using the Potential Series Expansion Method (PSEM), we conducted a stability analysis by examining the linearized state equations near these points. This analysis provides valuable insights into the dynamics and long-term behavior of asteroid (99942) Apophis. The eigenvalues of the equilibrium points, characterizing their stability properties, are presented in Table \ref{Eigenvalues_Apophis}. According to the topological classification by \citet{Jiang_2014} and \citet{Wang_2014}, equilibrium points $E_1$ and $E_3$ are unstable, as they have two real eigenvalues and four pure imaginary eigenvalues, making them Case 2, saddle-center-center. This configuration reveals two families of periodic orbits and a family of quasi-periodic orbits near each of these points. In contrast, equilibrium points $E_2$ and $E_4$ are linearly stable, falling under Case 1, with three pairs of pure imaginary eigenvalues, indicating three families of periodic orbits. \\

\begin{figure}[ht]
        \begin{center}
            $a_0 = 0.35$ km \hspace{4cm} $a_0 = 0.40$ km\\
            \includegraphics[width=0.48\linewidth]{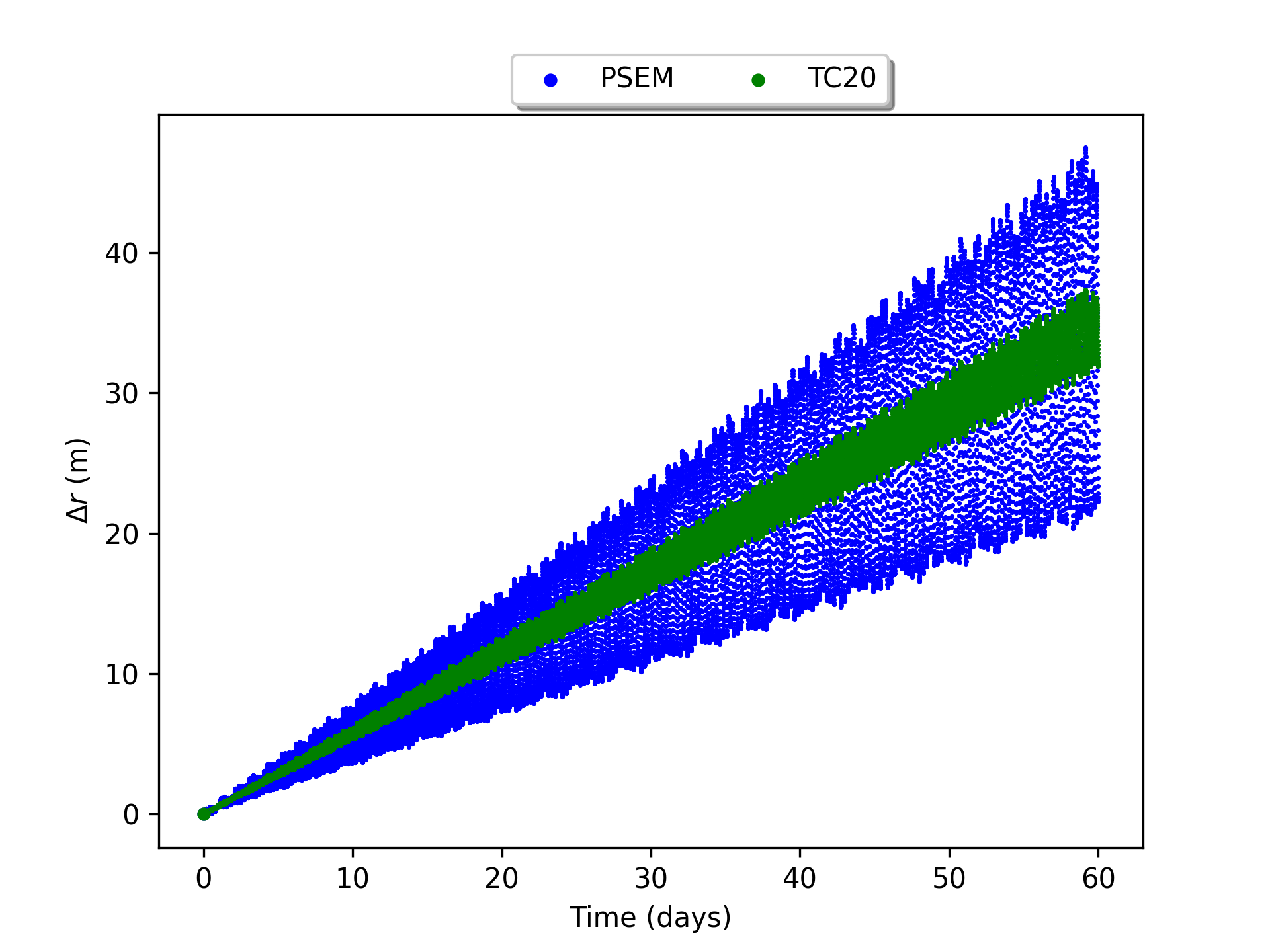}
            \includegraphics[width=0.48\linewidth]{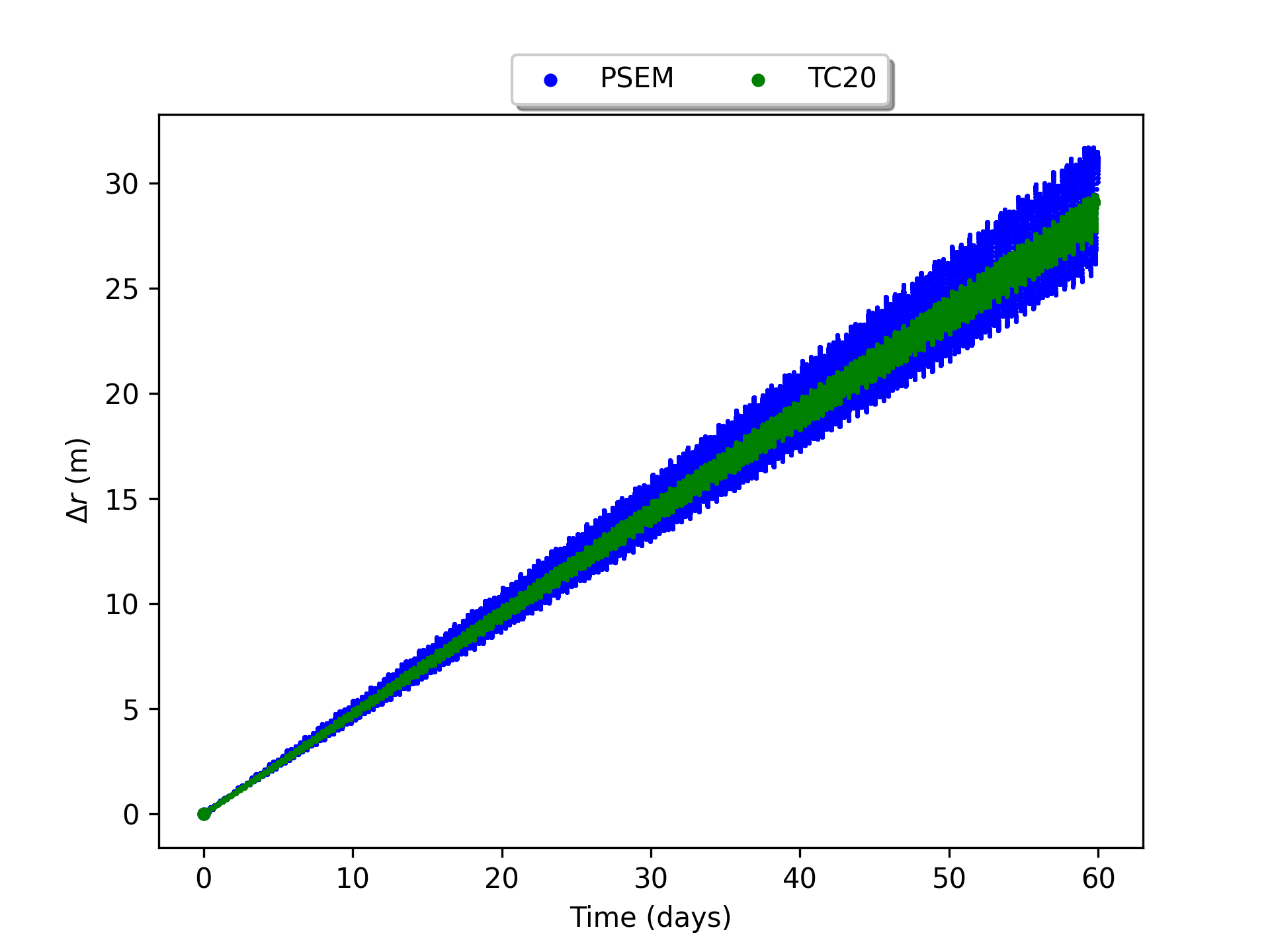}\\
            $a_0 = 0.50$ km \hspace{4cm} $a_0 = 0.60$ km\\
            \includegraphics[width=0.48\linewidth]{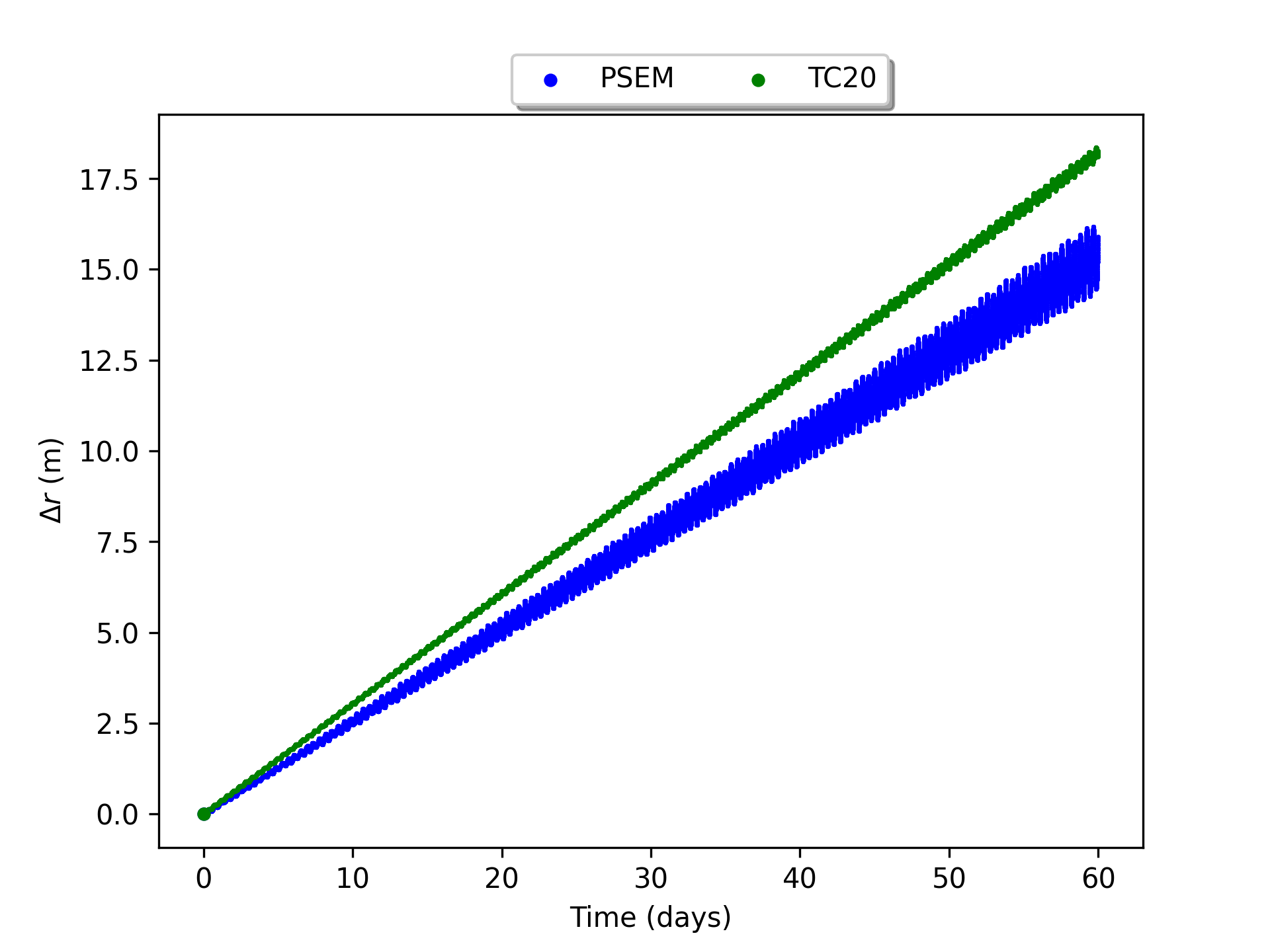}
            \includegraphics[width=0.48\linewidth]{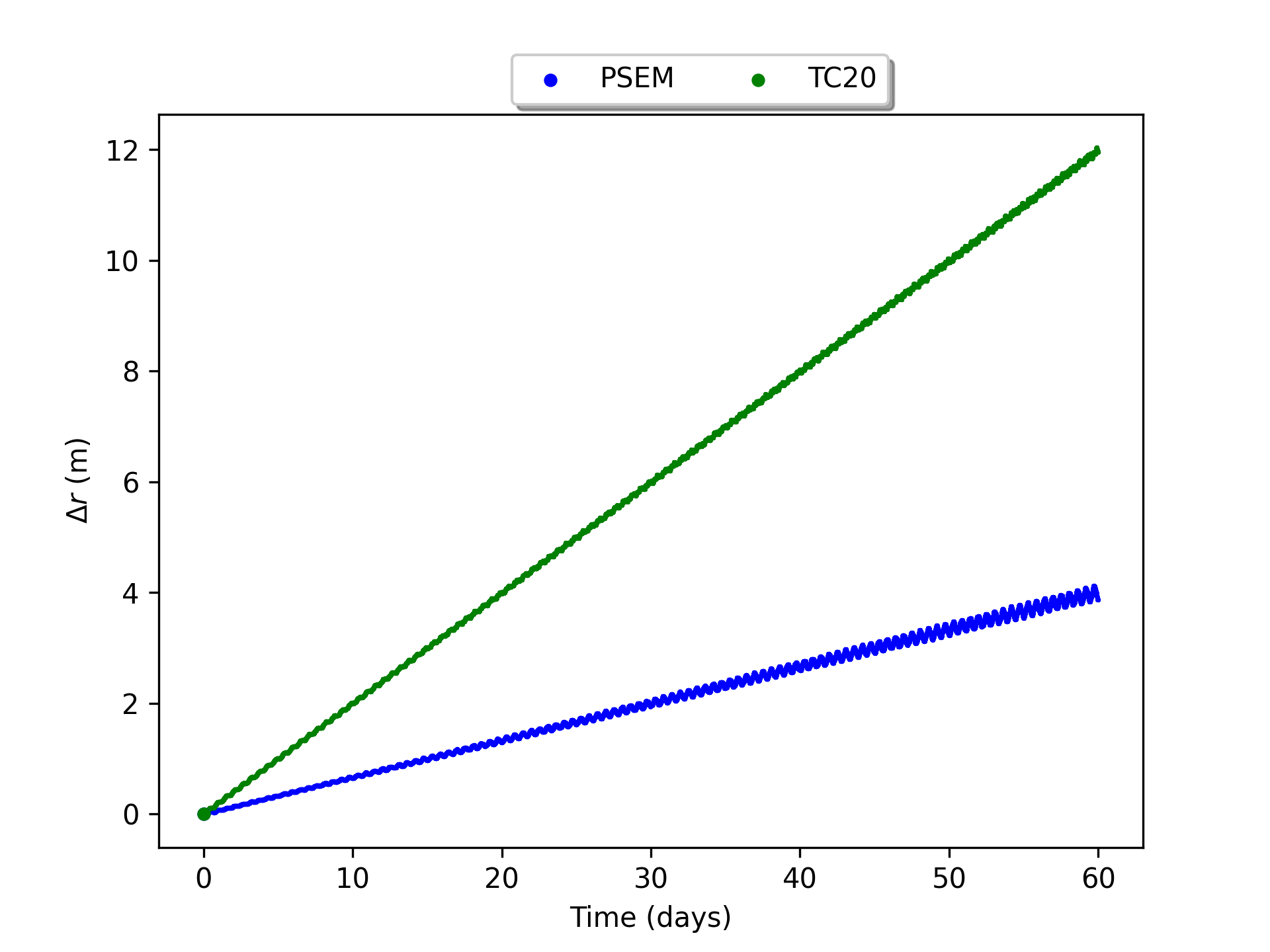}
        \end{center}
        \caption{Comparison of Orbit Differences Between PSEM, TC20, and Tsoulis Methods for a Spacecraft Orbiting Apophis} \label{fig_diff_orbit}
    \end{figure}

    When comparing the numerical integration times and precision of the three different methods presented in this work for calculating the orbit of a spacecraft around the asteroid Apophis, the Potential Series Expansion Method (PSEM, please see appendix 1) demonstrates significant advantages. The integration time using Tsoulis's method is 40 minutes and 3.378 seconds, while the Tetrahedron Center 20 (TC20) method requires 20 minutes and 10.404 seconds. In stark contrast, the PSEM completes the integration in just 2.682 seconds. Moreover, for orbits with semi-major axes greater than 0.5 km, the PSEM proves to be more precise than the TC20 method. Even for orbits closer than this limit, the precision of PSEM remains highly satisfactory. This is evident in Fig. \ref{fig_diff_orbit}, which shows the difference between the orbits calculated using PSEM or TC20 and those obtained with Tsoulis's method. The plot reveals that while the TC20 method aligns more closely with the Tsoulis method for closer orbits, PSEM's precision improves and surpasses TC20 as the semi-major axis increases, making it a highly efficient and accurate tool for modeling spacecraft trajectories around irregular celestial bodies.

\section{Conclusion}\

   The main objective of this work was to develop a method to model the external gravitational field of a small celestial body. Our model, called the series potential expansion method (PSEM), consists of applying the gravitational structure using a shaped polyhedral source, associated with the series potential expansion, similar to the spherical harmonics method. In this way, we developed a new approach to evaluate the external gravitational potential of a homogeneous polyhedron with constant density, presenting the advantage of explicitly displaying the gravitational field, facilitating algebraic manipulation to obtain the acceleration components and to calculate the speed increment, significantly reducing costs in terms of CPU time requirements while maintaining accuracy at a very acceptable level, as proven for the asteroids (87) Sylvia, (101955) Bennu, (99942) Apophis, and (25143) Itokawa. Furthermore, we determined equilibrium points, analyzed stability, investigated zero velocity planes, and calculated the relative errors between the gravitational field modeled by PSEM and the results obtained using both the classical polyhedral method of Tsoulis and Petrovic and the mass concentration method, demonstrating the computational efficiency of PSEM in modeling the gravitational potential of irregularly shaped bodies.
     
   In conclusion, the Potential Series Expansion Method (PSEM) not only maintains a high level of accuracy in modeling the gravitational potential of irregularly shaped celestial bodies but also reduces computational processing time by up to 98\% compared to the classical polyhedral method developed by Tsoulis, making it a highly efficient and effective tool for such analyses. It is important to note that the PSEM is valid and uniformly convergent outside the Brillouin sphere, the smallest sphere enclosing the entire mass of the body. In this region, the method provides reliable and accurate approximations of the gravitational field. However, as with spherical harmonic models, the series may diverge within the Brillouin sphere, limiting its applicability closer to the body's surface. Despite this limitation, the uniform convergence outside the sphere ensures its robustness for exterior applications, particularly in scenarios relevant to spacecraft dynamics and mission planning.

    \section*{Acknowledgements}
    The authors wish to express their appreciation for the support provided by grants 443116/2023-7 and 309089/2021-2 from the National Council for Scientific and Technological Development (CNPq); grant 88887.374148/2019-00 from the  Coordination for the Improvement of Higher Education Personnel (CAPES). Additionally, we would like to thank the Federal Institute of S\~{a}o Paulo for their support in facilitating the completion of this work.

    \section*{Author Contributions}
    M. L. Mota initiated the work and developed the mathematical framework. S. Aljbaae was responsible for the programming aspect, utilizing Python and the SymPy library to apply the mathematical components. A. F. B. A. Prado supervised and validation of the results. All authors reviewed and approved the final manuscript.
        
    \section*{Data availability} Any additional data not found in these sources can be obtained from the first author upon reasonable request.

    \begin{appendices}
        
        \section{Deduction of the integral of a polynomial function over the rectangular tetrahedron}\label{appendice_a}
            Due to the importance of this integral in deriving the gravitational potential model presented in this work, its derivation is provided below.
            
            The beta and gamma functions are defined by Equations (\ref{eq_B1}) and (\ref{eq_B2}), respectively:
             \begin{eqnarray}\label{eq_B1}
               \beta \left( {{n}_{1}}+1,{{n}_{2}}+1 \right)=\int_{0}^{1}{{{x}^{{{n}_{1}}}}{{\left( 1-x \right)}^{{{n}_{2}}}}dx}
                \end{eqnarray}\
                 \begin{eqnarray}\label{eq_B2}
               \Gamma \left( n+1 \right)=\int_{0}^{\infty }{{{e}^{-x}}{{x}^{n}}dx}=n!
                \end{eqnarray}\
                and are related through Equation (B.3):
                \begin{eqnarray}\label{eq_B3}
               \beta \left( m,n \right)=\frac{\Gamma \left( m \right)\Gamma \left( n \right)}{\Gamma \left( m+n \right)}=\frac{\left( m-1 \right)!\,\,\left( n-1 \right)!}{\left( m+n-1 \right)!}
                \end{eqnarray}\
            Therefore, the integral of the function $f\left( x,y,z \right)={{x}^{{{n}_{1}}}}{{y}^{{{n}_{2}}}}{{z}^{{{n}_{3}}}}$ over the rectangular tetrahedron $W$, with vertices located at points \(O\left( {0,0,0} \right)\), \({V_1}\left( {1,0,0} \right)\), \({V_2}\left( {0,1,0} \right)\) and \({V_3}\left( {0,0,1} \right)\), is expressed by Equation (\ref{eq_B4}):
            
           \begin{align}\label{eq_B4}
              & I=\iiint\limits_{W}{{{x}^{{{n}_{1}}}}{{y}^{{{n}_{2}}}}{{z}^{{{n}_{3}}}}dx\,dy\,dz=\int_{0}^{1}{\int_{0}^{1-z}{\int_{0}^{1-z-y}{{{x}^{{{n}_{1}}}}{{y}^{{{n}_{2}}}}{{z}^{{{n}_{3}}}}dx\,dy\,dz}}}} \nonumber \\ 
             & \,\,\,\,\,\,\,\,\,\,\,\,\,\,\,\,\,\,\,\,\,\,\,\,\,\,\,\,\,\,\,\,\,\,\,\,\,\,\,\,\,\,\,\,\,\,\,\,\,\,=\frac{1}{{{n}_{1}}+1}\int_{0}^{1}{\int_{0}^{1-z}{{{\left( 1-z-y \right)}^{{{n}_{1}}+1}}{{y}^{{{n}_{2}}}}{{z}^{{{n}_{3}}}}dy\,dz}} \\ 
             & \,\,\,\,\,\,\,\,\,\,\,\,\,\,\,\,\,\,\,\,\,\,\,\,\,\,\,\,\,\,\,\,\,\,\,\,\,\,\,\,\,\,\,\,\,\,\,\,\,\,=\frac{1}{{{n}_{1}}+1}\int_{0}^{1}{\int_{0}^{1-z}{{{\left( 1-z \right)}^{{{n}_{1}}+1}}{{\left( 1-\frac{y}{1-z} \right)}^{{{n}_{1}}+1}}{{y}^{{{n}_{2}}}}{{z}^{{{n}_{3}}}}dy\,dz}} \nonumber \nonumber \ 
            \end{align}
            Defining $Y=\frac{y}{1-z}$, then, $y=\left( 1-z \right)Y$. Therefore, $\frac{dy}{dY}=\left( 1-z \right)$, which, substituted into Equation (\ref{eq_B4}), produces Equation (\ref{eq_B5}):
            
            \begin{eqnarray}\label{eq_B5}
               I=\frac{1}{{{n}_{1}}+1}\int_{0}^{1}{\int_{0}^{1}{{{\left( 1-z \right)}^{{{n}_{1}}+1}}{{\left( 1-Y \right)}^{{{n}_{1}}+1}}{{\left( 1-z \right)}^{{{n}_{2}}}}{{Y}^{{{n}_{2}}}}{{z}^{{{n}_{3}}}}\left( 1-z \right)dY\,dz}}
                \end{eqnarray}\
            By algebraically manipulating Equation (\ref{eq_B5}), we arrive at Equation (\ref{eq_B6}):
            \begin{eqnarray}\label{eq_B6}
               I=\frac{1}{{{n}_{1}}+1}\int_{0}^{1}{\int_{0}^{1}{{{\left( 1-Y \right)}^{{{n}_{1}}+1}}{{Y}^{{{n}_{2}}}}{{\left( 1-z \right)}^{\left( {{n}_{1}}+{{n}_{2}}+2 \right)}}{{z}^{{{n}_{3}}}}dY\,dz}}
                \end{eqnarray}\
            which can be rewritten using Equation (\ref{eq_B7}):
            \begin{eqnarray}\label{eq_B7}
              I=\frac{1}{{{n}_{1}}+1}\int_{0}^{1}{\beta \left( {{n}_{2}}+1,{{n}_{1}}+2 \right){{\left( 1-z \right)}^{\left( {{n}_{1}}+{{n}_{2}}+2 \right)}}{{z}^{{{n}_{3}}}}dz}
                \end{eqnarray}\
             \begin{eqnarray}\label{eq_B8}
             I=\frac{\beta \left( {{n}_{2}}+1,{{n}_{1}}+2 \right)}{{{n}_{1}}+1}\int_{0}^{1}{{{\left( 1-z \right)}^{\left( {{n}_{1}}+{{n}_{2}}+2 \right)}}{{z}^{{{n}_{3}}}}dz}
                \end{eqnarray}\   
            In turn, Equation (\ref{eq_B8}) subsequently produces Equations (\ref{eq_B9}) and (\ref{eq_B10}):
            \begin{eqnarray}\label{eq_B9}
              I=\frac{\beta \left( {{n}_{2}}+1,{{n}_{1}}+2 \right)}{{{n}_{1}}+1}\beta \left( {{n}_{3}}+1,{{n}_{1}}+{{n}_{2}}+3 \right)
                \end{eqnarray}\
            
            \begin{eqnarray}\label{eq_B10}
              I=\frac{1}{\left( {{n}_{1}}+1 \right)}\frac{{{n}_{2}}!\left( {{n}_{1}}+1 \right)!}{\left( {{n}_{1}}+{{n}_{2}}+2 \right)!}\frac{{{n}_{3}}!\left( {{n}_{1}}+{{n}_{2}}+2 \right)!}{\left( {{n}_{1}}+{{n}_{2}}+{{n}_{3}}+3 \right)!}
                \end{eqnarray}\
            Therefore, the integral of the polynomial function $f\left( x,y,z \right)={{x}^{{{n}_{1}}}}{{y}^{{{n}_{2}}}}{{z}^{{{n}_{3}}}}$ over the rectangular tetrahedron $W$, is given by Equation (\ref{eq_B11})
            \begin{eqnarray}\label{eq_B11}
              I=\iiint\limits_{W}{{{x}^{{{n}_{1}}}}{{y}^{{{n}_{2}}}}{{z}^{{{n}_{3}}}}dx\,dy\,dz=\frac{{{n}_{1}}!\,\,{{n}_{2}}!\,\,{{n}_{3}}!}{\left( {{n}_{1}}+{{n}_{2}}+{{n}_{3}}+3 \right)!}}.
            \end{eqnarray}\\

    \section{Calculation of Potentials for Asteroid Apophis}\label{appendice_b}
            Regarding the calculations of potentials applying our method, as an example, we chose the asteroid Apophis. Using the mass \( M = 5.31 \times 10^{10} \, \text{kg} \) and the gravitational constant \( G = 6.67430 \times 10^{-11} \, \text{m}^3 \text{kg}^{-1} \text{s}^{-2} \), we present below some of its potentials obtained analytically:
            \begin{align}\label{eq_combined}
            U_0 &= \frac{3.543947 \times 10^{-9}}{\sqrt{x^2 + y^2 + z^2}} \nonumber \\
            U_1 &= \frac{2.101984 \times 10^{-22} x}{(x^2 + y^2 + z^2)^{3/2}} + \frac{1.863005 \times 10^{-22} y}{(x^2 + y^2 + z^2)^{3/2}} - \frac{2.145309 \times 10^{-22} z}{(x^2 + y^2 + z^2)^{3/2}} \nonumber \\
            U_2 &= \frac{1.569552 \times 10^{-11} x^2}{(x^2 + y^2 + z^2)^{5/2}} + \frac{7.060517 \times 10^{-23} xy}{(x^2 + y^2 + z^2)^{5/2}} - \frac{5.303897 \times 10^{-12} y^2}{(x^2 + y^2 + z^2)^{5/2}} \nonumber \\
            &\quad - \frac{6.817990 \times 10^{-23} xz}{(x^2 + y^2 + z^2)^{5/2}} + \frac{4.610980 \times 10^{-23} yz}{(x^2 + y^2 + z^2)^{5/2}} - \frac{1.039163 \times 10^{-11} z^2}{(x^2 + y^2 + z^2)^{5/2}} \nonumber \\
            U_3 &= -\frac{2.013911 \times 10^{-12} x^3}{(x^2 + y^2 + z^2)^{7/2}} + \frac{3.005219 \times 10^{-13} x^2 y}{(x^2 + y^2 + z^2)^{7/2}} + \frac{4.941320 \times 10^{-12} x y^2}{(x^2 + y^2 + z^2)^{7/2}} \nonumber \\
            &\quad + \frac{1.524291 \times 10^{-14} y^3}{(x^2 + y^2 + z^2)^{7/2}} - \frac{5.070411 \times 10^{-12} x^2 z}{(x^2 + y^2 + z^2)^{7/2}} - \frac{2.442363 \times 10^{-12} xyz}{(x^2 + y^2 + z^2)^{7/2}} \nonumber \\
            &\quad + \frac{1.614581 \times 10^{-12} y^2 z}{(x^2 + y^2 + z^2)^{7/2}} + \frac{1.100414 \times 10^{-12} x z^2}{(x^2 + y^2 + z^2)^{7/2}} - \frac{3.462506 \times 10^{-13} y z^2}{(x^2 + y^2 + z^2)^{7/2}} \nonumber \\
            &\quad + \frac{1.151943 \times 10^{-12} z^3}{(x^2 + y^2 + z^2)^{7/2}}
            \end{align}
            where \( U_0 \), \( U_1 \), \( U_2 \), and \( U_3 \) are the potentials of degree 0, 1, 2, and 3, respectively, related to the asteroid Apophis. It can be observed that, due to the fact that the center of mass coincides with the origin of the coordinate system fixed to the body, \( U_1 = 0 \). Additionally, because the axes of the coordinate system fixed to the body coincide with the main axes of inertia, the coefficients of \( xy \), \( xz \), and \( yz \) in the potential \( U_2 \) are zero. In our study on asteroid Apophis, we developed the potential up to degree 11, achieving excellent accuracy and reduced computational cost in calculating the potential, determining the equilibrium points, and studying stability and orbital simulations.

            This detailed analytical approach was also applied similarly to other asteroids, such as (87) Sylvia, (101955) Bennu, and (25143) Itokawa, showing consistent results.

    \end{appendices}

\bibliography{sn-article_v3.bib}

\end{document}